\documentclass[12pt]{article}
\usepackage{amsmath}
\usepackage{multirow}
\usepackage{amsfonts}
\usepackage{amssymb,graphics,psfrag,float}
\usepackage{array,epsfig,multirow,graphicx}
\usepackage{comment}
\usepackage{slashed}
\usepackage{hyperref}
\usepackage{tensor}
\usepackage{booktabs}
\usepackage{colortbl}
\usepackage{hhline}
\usepackage{mathtools}
\usepackage{enumitem}
\usepackage{xfrac}
\usepackage{tikz}
\usepackage[nosort]{cite}
\usetikzlibrary{decorations.markings}
\usetikzlibrary{shapes,arrows}
\usetikzlibrary{decorations.pathreplacing}
\usetikzlibrary{arrows,positioning}

\def\hybrid{\topmargin -20pt    \oddsidemargin 0pt
        \headheight 0pt \headsep 0pt 
        \textwidth 6.25in      
        \textheight 9 in      
        \marginparwidth .875in
        \parskip 5pt plus 1pt
          \jot = 1.5ex
  }
\hybrid
\numberwithin{equation}{section}
\numberwithin{table}{section}

\newcommand{\beq}{\begin{equation}}
\newcommand{\eeq}{\end{equation}}
\newcommand{\be}{\begin{equation}}
\newcommand{\ee}{\end{equation}}
\newcommand{\bea}{\begin{eqnarray}}
\newcommand{\eea}{\end{eqnarray}}
\newcommand{\ben}{\begin{eqnarray*}}
\newcommand{\een}{\end{eqnarray*}}               
\newcommand{\ba}{\begin{aligned}}
\newcommand{\ea}{\end{aligned}}
\newcommand{\bt}{\begin{tabular}}
\newcommand{\et}{\end{tabular}}
\newcommand{\bc}{\begin{center}}
\newcommand{\ec}{\end{center}}

\newcommand{\cE}{\mathcal{E}}

\newcommand{\cC}{\mathcal{C}}
\newcommand{\cD}{\mathcal{D}}

\newcommand{\cK}{\mathcal{K}}
\newcommand{\cN}{\mathcal{N}}

\newcommand{\cI}{\mathcal{I}}
\newcommand{\cJ}{\mathcal{J}}

\newcommand{\cM}{\mathcal M}

\DeclareMathOperator{\rk}{rank}

\newcommand{\bbZ}{\mathbb{Z}}

\newcommand{\nn}{\nonumber}

\newcommand{\cref}{{\bf [check ref]}}

\newcommand{\tr}{\mathrm{tr}}

\newcommand{\dynkinradius}{.05cm}
\newcommand{\dynkinstep}{.25cm}
\newcommand{\dynkindot}[2]{\fill (\dynkinstep*#1,\dynkinstep*#2) circle (\dynkinradius); }

\newcommand{\dynkinline}[4]{\draw[thin] (\dynkinstep*#1,\dynkinstep*#2) -- (\dynkinstep*#3,\dynkinstep*#4);}
\newcommand{\dynkindots}[4]{\draw[dotted] (\dynkinstep*#1,\dynkinstep*#2) -- (\dynkinstep*#3,\dynkinstep*#4);}
\newcommand{\dynkindoubleline}[4]{\draw[double,postaction={decorate}] (\dynkinstep*#1,\dynkinstep*#2) -- (\dynkinstep*#3,\dynkinstep*#4);}
\newcommand{\dynkintripleline}[4]{\draw[double,postaction={decorate}] (\dynkinstep*#1,\dynkinstep*#2) -- (\dynkinstep*#3,\dynkinstep*#4);
				  \draw (\dynkinstep*#1,\dynkinstep*#2) -- (\dynkinstep*#3,\dynkinstep*#4);}

\newenvironment{dynkin}{\begin{tikzpicture}[decoration={markings,mark=at position 0.6 with {\arrow[line width=0.15mm]{>}}}]}
{\end{tikzpicture}}
\definecolor{mppgreen}{RGB}{17,102,86}
\definecolor{mppgray}{RGB}{221,222,214}

\def\blfootnote{\xdef\@thefnmark{}\@footnotetext}
\long\def\symbolfootnote[#1]#2{\begingroup%
\def\thefootnote{\fnsymbol{footnote}}\footnote[#1]{#2}\endgroup}

\begin{document}

\baselineskip=15pt

\begin{titlepage}
\begin{flushright}
\parbox[t]{1.8in}{\begin{flushright}
CERN-PH-TH-2015-230\\
MPP-2015-203 \end{flushright}}
\end{flushright}

\begin{center}

\vspace*{ 1.2cm}

{\Large \bf  The Arithmetic of Elliptic Fibrations\\[.2cm]
                   in Gauge Theories on a Circle}

\vskip 1.2cm

\renewcommand{\thefootnote}{}
\begin{center}
 {Thomas W.~Grimm$^{\,1,2}$, Andreas Kapfer$^{\, 1}$ and Denis Klevers$^{\, 3}$\ \footnote{grimm,\ kapfer@mpp.mpg.de,\ denis.klevers@cern.ch}}
\end{center}
\vskip .2cm
\renewcommand{\thefootnote}{\arabic{footnote}}

{$^1$ Max-Planck-Institut f\"ur Physik, \\
F\"ohringer Ring 6, 80805 Munich, Germany

$^2$ Institute for Theoretical Physics and \\
Center for Extreme Matter and Emergent Phenomena,\\
Utrecht University, Leuvenlaan 4, 3584 CE Utrecht, The Netherlands

$^3$ Theory Group, Physics Department, CERN, \\ CH-1211, Geneva 23, Switzerland}

 \vspace*{.2cm}

\end{center}

 \renewcommand{\thefootnote}{\arabic{footnote}}
 
\begin{center} {\bf ABSTRACT } \end{center}

The geometry of elliptic fibrations translates to the physics of gauge theories in F-theory. 
We systematically develop the dictionary between arithmetic structures on elliptic curves as well 
as desingularized elliptic fibrations and symmetries of gauge theories on a 
circle. We show that the Mordell-Weil group law matches integral 
large gauge transformations around the circle in Abelian gauge theories and explain 
the significance of Mordell-Weil torsion in this context. 
We also use Higgs transitions and circle large gauge transformations to 
introduce a group law for genus-one fibrations with multi-sections. 
Finally, we introduce a novel arithmetic structure on elliptic fibrations with 
non-Abelian gauge groups in F-theory. It is defined on the set of exceptional 
divisors resolving the singularities and divisor classes of sections of the fibration.
This group structure can be matched with certain integral non-Abelian large gauge transformations around the circle
when studying the theory on the lower-dimensional Coulomb branch. 
Its existence is required by consistency with Higgs transitions 
from the non-Abelian theory to its Abelian phases in which it becomes 
the Mordell-Weil group. 
This hints towards the existence of a new
underlying geometric symmetry.

\hfill {October, 2015}
\end{titlepage}

\tableofcontents

\newpage



\section{Introduction}

In recent years the connection of the geometry of elliptic curves to 
gauge theories in various dimensions has been explored intensively 
by using F-theory \cite{Vafa:1996xn}. In F-theory two auxiliary dimensions
need to be placed on a two-torus whose complex structure 
is identified with the Type IIB dilaton-axion.  Its variations are then encoded by 
the geometry of a two-torus fibration in F-theory. Magnetic sources for the 
dilaton-axion are seven-branes that support gauge theories. Several features 
of these gauge theories can thus be studied using two-torus  fibrations.
At first, the F-theory approach seems to suggest 
that the connection between geometry and gauge theories is rather direct. 
However, it turns out that the geometry of elliptic fibrations should 
rather be related to gauge theories compactified on a circle. 
This can be understood by realizing that the volume of the two-torus 
is unphysical in F-theory and that there is no notion of an actual 
twelve-dimensional background geometry. The geometry of the 
elliptic fibration of F-theory is only fully probed in the dual M-theory compactification.
M-theory compactified on an elliptic fibration yields the effective theory of F-theory 
compactified on an additional circle. In particular, one is therefore forced to 
relate the geometry of elliptic fibrations with gauge theories on a circle. 
In this work we will further develop 
the dictionary between the geometry of elliptic fibrations and 
matter-coupled gauge theories on a circle. Our focus here will be 
on revealing geometric symmetry transformations that correspond to
large gauge transformations around the circle. 

As a first example of such a relation we will study Abelian gauge theories 
on a circle. In an F-theory compactification the number of massless 
$U(1)$ fields can be related to the number of rational sections or multi-sections
minus one \cite{Morrison:1996pp}. 
The  section that is not counted here has to be identified with 
the Kaluza-Klein vector  
obtained from the higher-dimensional metric when placing the gauge theory 
on the additional circle. Recent 
progress in understanding $U(1)$ gauge groups in F-theory can be found, for 
example, in the references \cite{Grimm:2010ez,Park:2011ji,Morrison:2012ei,Braun:2013yti,Borchmann:2013jwa,Cvetic:2013nia,Grimm:2013oga,Braun:2013nqa,Cvetic:2013uta,Borchmann:2013hta,Cvetic:2013jta,Cvetic:2013qsa,Klevers:2014bqa,Lawrie:2015hia,Cvetic:2015ioa}. The fact that smooth geometries carry information 
about a circle reduced theory becomes particularly apparent 
in models with rational sections in which the mass hierarchy between 
Kaluza-Klein masses and lower-dimensional Coulomb branch masses 
is non-trivial \cite{Grimm:2013oga}. 
In other words, it was key in \cite{Grimm:2013oga,Grimm:2015zea} that despite 
the fact that massive states have to be integrated out in the circle 
compactified effective theory some cutoff independent information about the massive tower 
has to be kept. 
In particular, the one-loop Chern-Simons terms carry information 
about the representations of the higher-dimensional chiral 
spectrum supplemented with a table of signs for each state \cite{Grimm:2011fx,Cvetic:2012xn}.
This extra information can be summarized in so-called box graphs introduced 
in \cite{Hayashi:2014kca,Braun:2014kla}, see also \cite{Esole:2014bka,Esole:2014hya}. 
In general, however, it is important to also keep track of an 
integer label for each dimensionally reduced state that encodes the 
mass hierarchy between the Kaluza-Klein mass and the Coulomb branch mass. 
For models with only a multi-section, see 
\cite{deBoer:2001px,Braun:2014oya,Morrison:2014era,Klevers:2014bqa,Cvetic:2015moa,Braun:2014qka} for representative works, 
the requirement of a lower-dimensional approach is even more pressing. As discussed in 
\cite{Anderson:2014yva,Garcia-Etxebarria:2014qua,Mayrhofer:2014haa,Mayrhofer:2014laa,Cvetic:2015moa} the multi-section should be understood as a mixing of 
the higher-dimensional $U(1)$s and the Kaluza-Klein vector.  

The first goal of this work is to formalize the relationship between 
geometries with rational sections and circle reduced gauge theories further.
We carefully identify the Mordell-Weil group acting on rational 
sections as large gauge transformations along 
the circle, as already suggested in \cite{Grimm:2015zea}. The Mordell-Weil group is a discrete 
finitely-generated Abelian group that captures key information about the arithmetic 
of elliptic fibrations. We show that there is indeed a one-to-one correspondence of 
large gauge transformations with generators of the Mordell-Weil group. 
The free part of the Mordell-Weil group is identified with Abelian large gauge 
transformations, while its torsion part is related to special fractional 
non-Abelian large gauge transformations. As a byproduct we explore the 
geometric relationship between the existence of fractional Abelian charges 
of matter states and the presence of a non-Abelian gauge group. 

A second goal of this work is to use our understanding of the arithmetic
for geometries with rational sections to provide evidence for the existence of a natural group law acting on
fibrations with multi-sections. We call this group \textit{extended Mordell-Weil group}, despite the fact 
that there is formally no Mordell-Weil group for multi-sections. We also define 
a \textit{generalized Shioda map} that allows to explore the physical implications of the
group action. 
Furthermore, we rigorously establish the correspondence of the proposed group 
action on the divisor level with large gauge transformations around the circle. In many 
examples it is known that there exist geometric transitions from a model 
with several sections to a model with multi-section \cite{Morrison:2014era,Anderson:2014yva,Klevers:2014bqa,Garcia-Etxebarria:2014qua,Mayrhofer:2014haa,Mayrhofer:2014laa,Cvetic:2015moa}. Physically 
this corresponds to a Higgsing of charged matter states. By construction 
the group law of the extended Mordell-Weil group should be inherited from the 
setup with multiple sections. Accordingly, it 
trivially reduces to the standard Mordell-Weil group law in the presence 
of sections only. As first explained in \cite{Grimm:2015zea}, in geometries 
with multiple sections the fact that one is free to 
choose a zero-section in the geometry to perform the F-theory limit is 
deeply linked to anomaly freedom of the higher-dimensional Abelian 
gauge theory. Using the extended Mordell-Weil group one would be
able to extend these arguments to this larger class of geometries and their 
corresponding effective theories.

The third goal of this paper it to extend the discussion to fully include 
matter-coupled non-Abelian gauge theories with gauge group $G$.
Placing these theories on a circle we perform 
large gauge transformations along the circle and explore 
the associated arithmetic structure in the geometry. More precisely, we 
are interested in examining the impact of non-Abelian 
large gauge transformation on the $U(1)^{r+1}$ gauge theory 
obtained in the lower-dimensional Coulomb branch. Here $r$ is 
the rank of $G$ and the additional $U(1)$ is the Kaluza-Klein vector 
of the higher-dimensional metric. The circle gauge transformations 
mix these vectors non-trivially and it was argued in \cite{Grimm:2015zea} that 
the identification of the effective theories is only possible if 
non-Abelian anomalies are cancelled. However, the group structure encoding 
the allowed changes remained mysterious in \cite{Grimm:2015zea}. In this work 
we show that there indeed is a natural group structure on the set of exceptional 
divisors and rational sections corresponding to these large gauge transformations. 
We also make progress in identifying the geometric symmetry 
corresponding to such transformations. 
First, we show that the transformed exceptional divisors and rational sections can have a standard geometry
interpretation. Second, we employ that  a non-Abelian gauge theory with adjoints is related
to an Abelian theory with  $r$  $U(1)$s by Higgsing/unHiggsing corresponding to complex structure deformations in
the geometry. Using recent results in \cite{Morrison:2014era,Cvetic:2015ioa} we show
that the postulated group structure on the exceptional divisors gets mapped precisely to 
the  usual Mordell-Weil group law of the geometry corresponding to the Abelian theory. 
All this hints at the existence of a geometrical symmetry directly in the geometry associated to the non-Abelian theory that is yet to be formulated explicitly. In fact, on the field theory level this is clear since the circle-compactified theories only differ by a 
non-Abelian large gauge transformation. 
This is obviously a symmetry of an anomaly free gauge theory, which indicates that 
the corresponding geometry should be considered physically equivalent to the 
original elliptic fibration. 

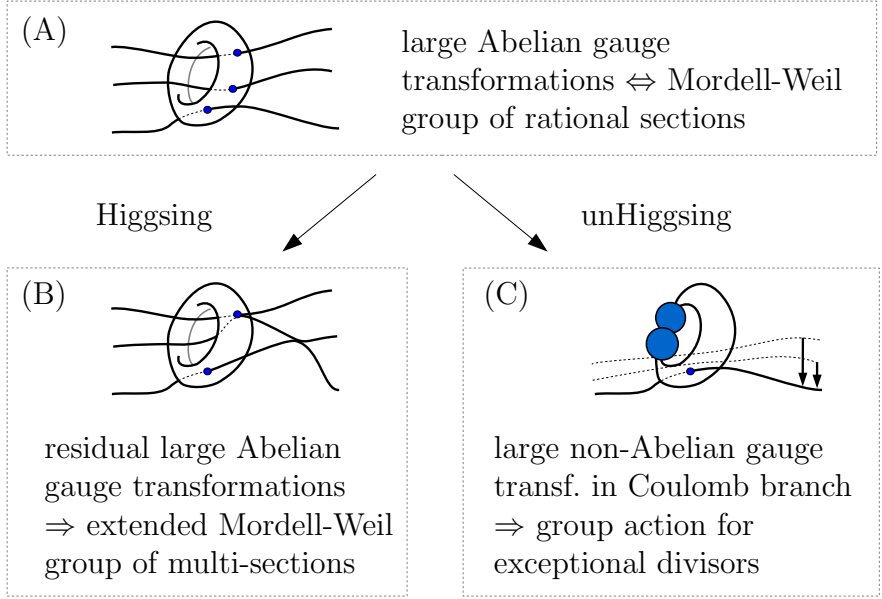
\begin{figure}[t]
\begin{center}
\includegraphics[scale=0.6]{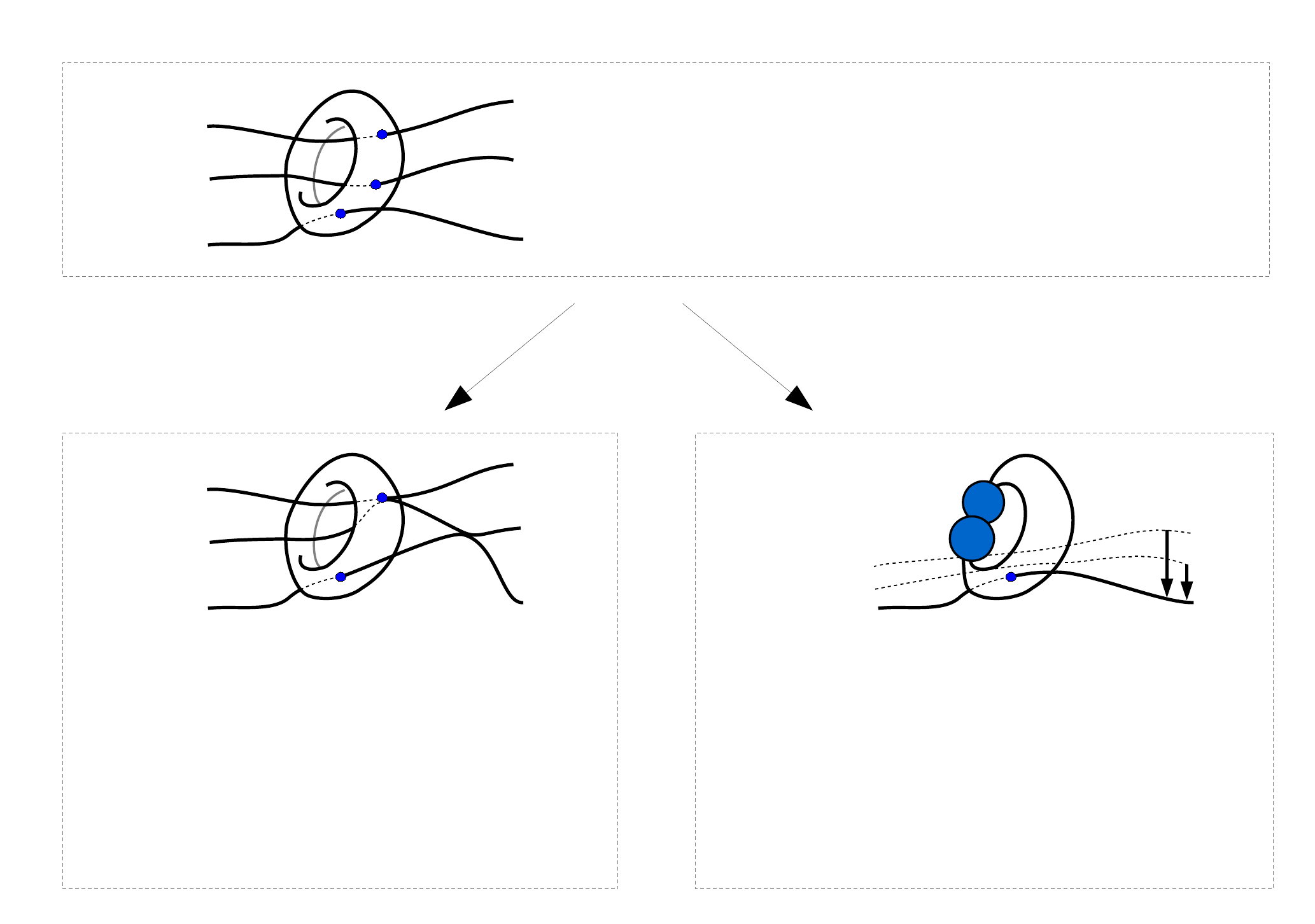}
\begin{picture}(0,0)
\put(-200,195){
\begin{minipage}{6cm} large Abelian gauge \\
transformations  $\Leftrightarrow$ Mordell-Weil   \\
group of rational sections   \\
\end{minipage}}
\put(-335,40){
\begin{minipage}{6cm} residual large Abelian\\
gauge transformations\\
$\Rightarrow$ extended Mordell-Weil\\
group of multi-sections 
\end{minipage}}
\put(-165,40){
\begin{minipage}{6cm} large non-Abelian gauge\\
 transf.~in Coulomb branch\\
$\Rightarrow$ group action for\\
exceptional divisors
\end{minipage}}
\put(-340,220){(A)}
\put(-340,120){(B)}
\put(-165,120){(C)}
\put(-375,150){\begin{minipage}{6cm} \begin{center} 
Higgsing \end{center} \end{minipage}}
\put(-185,150){\begin{minipage}{6cm} \begin{center} 
unHiggsing \end{center} \end{minipage}}
\end{picture}
\vspace*{-.2cm}

\begin{minipage}{12cm}
\caption{Schematic depiction of the various geometric configurations considered in 
this work. The geometries are related by geometric transitions describing 
Higgsing and unHiggsing processes.    \label{HuH-systematics}}
\end{minipage}
\vspace{-1cm}
\end{center}
\end{figure}

Before starting with the actual computations, let us 
note that our intuition can be summarized by using \autoref{HuH-systematics}
as follows. First, we are able 
to establish the relation of the Mordell-Weil group to large Abelian gauge transformations
in elliptic fibrations with rational sections depicted in (A).
Using a geometric transition, which corresponds to Higgsing in field theory, 
the resulting fibrations might only admit multi-sections, see (B). 
Therefore, one expects an extended Mordell-Weil group structure for such geometries.
Furthermore, a geometry with rational sections might arise via a geometric 
transition describing a non-Abelian gauge theory. Again this is described by 
a Higgsing in field theory. Such transitions motivate us to transfer the 
Mordell-Weil group structure to a geometry with exceptional divisors. The 
resulting group structure corresponds to large non-Abelian gauge transformation for an 
F-theory effective field theory compactified on a circle and pushed to the Coulomb branch.

The paper is organized as follows. In \autoref{sec:gauge_circle}
we discuss the compactification of Abelian and non-Abelian 
gauge theories on a circle. We introduce our notation and the 
relevant large gauge transformations along the circle. 
The treatment will be valid both for six- and four-dimensional gauge theories. We also
briefly comment on anomalies and one-loop Chern-Simons terms induced after circle 
compactification. A summary of some basic facts about F-theory compactifications 
and elliptic fibrations is provided in \autoref{Ftheoryfacts}.
In \autoref{sec:ec_group_structures} we focus on geometries with rational 
sections and the corresponding Abelian parts of the gauge theory. The Mordell-Weil group action is mapped to large gauge 
transformations along the circle.
We also discuss the impact of torsion from this perspective.
In \autoref{sec:multi_group} we then turn to geometries with multi-sections. 
Insights obtained by using Higgs transitions allow us to define an extended 
Mordell-Weil group of multi-sections and a generalized Shioda map.  
In \autoref{Arithmetics_nonAb} we extend the analysis further 
to cover non-Abelian gauge groups. We argue for the existence of 
a group law on the exceptional divisors and rational sections that is
shown to be induced by large gauge transformations of the Cartan gauge fields. 
We geometrically motivate its existence 
by explicitly considering Higgsings to Abelian gauge theories.

\section{Symmetries of gauge theories on the circle} \label{sec:gauge_circle}

In this section we discuss a number of symmetries that 
are encountered when considering six- and four-dimensional 
gauge theories on a circle. In \autoref{sec:gensetup+reduction} we first 
introduce our notation and summarize some basic 
facts about circle compactifications of gauge theories. 
In \autoref{sec:lgts_anomalies} we study the manifestation of large gauge 
transformations with gauge parameters supported on the circle. 
We comment on the rearrangement of the Kaluza-Klein spectrum
and the significance of one-loop induced Chern-Simons terms on the Coulomb branch of the gauge theory.

\subsection{General setup and circle reduction} \label{sec:gensetup+reduction}

We begin by introducing some general notions about Abelian and 
non-Abelian gauge theories in six and four space-time dimensions. 
Let us denote by $G$ a simple gauge group with gauge bosons $\hat A$. Introducing 
the Lie algebra generators $T_\cI$, $\cI = 1,\ldots,\text{dim}\,G$ we expand 
\beq \label{hatA_expand}
   \hat A = \hat A^\cI T_\cI  = \hat A^I T_I + \hat A^{\boldsymbol{\alpha}} T_{\boldsymbol{\alpha} }
\eeq
where $T_I$, $I = 1 ,\dots ,\rk G$ are the generators of the Cartan subalgebra and $T_{\boldsymbol \alpha}$
are the remaining generators labeled by the roots ${\boldsymbol \alpha}$. 
In addition, we will allow for a number $n_{U(1)}$ of Abelian gauge bosons that are 
denoted by $\hat A^m$ with $m =1,\dots ,n_{U(1)}$.

In the next step we compactify this theory on a circle and push it to the Coulomb branch
by allowing for a non-vanishing Wilson line background of the gauge fields reduced on the 
circle. From the four- or six-dimensional metric one finds at lowest level the three-
or five-dimensional metric 
$g_{\mu \nu}$, the Kaluza-Klein vector $A^0$, and the radius $r$ of the circle.
Thus, the higher-dimensional line element is expanded as
\begin{align}\label{e:metric_reduction}
 d \hat s^2 = g_{\mu\nu}dx^\mu dx^\nu + r^2 Dy^2 \, , \qquad Dy := dy - A^0_\mu dx^\mu \, ,
\end{align}
with $x^\mu$ the three- or five-dimensional coordinates and $y$ the coordinate along the circle.
Performing the Kaluza-Klein reduction of the vector fields one finds at lowest Kaluza-Klein level  
a number of $\text{dim}\, G$ gauge fields $A^\cI$ and
$\text{dim}\, G$ Wilson line scalars $\zeta^\cI$ from reducing $\hat A$. In addition 
one has  $n_{U(1)}$ $U(1)$ gauge fields $A^m$ and Wilson line scalars $\zeta^m$ 
from reducing $\hat A^m$.
In particular the gauge fields are expanded as
\begin{align} \label{red-AI}
 \hat A^\cI = A^\cI - \zeta^\cI r Dy\  , \qquad \hat A^m = A^m - \zeta^m r Dy \ .
\end{align}
The $A^\cI$ are now gauge fields of the lower-dimensional version of the gauge group $G$, 
while the $\zeta^\cI$ transform in the adjoint representation of $G$. In other words, denoting the gauge parameters by $\Lambda^\cI(x)$ and $\Lambda^m(x)$, 
   one has  
\bea \label{lower-dim-gauge}
   \delta A^\cI &=&  d\Lambda^\cI + f^\cI_{\cJ \cK} \Lambda^\cJ   A^\cK \ , 
   \qquad \  \ \delta \zeta^\cI  =  f^{\cI}_{\cJ \cK} \zeta^\cJ \Lambda^\cK\ ,\\
  \delta A^m &=&  d\Lambda^m  \ , 
   \qquad \qquad \qquad  \qquad \quad \, \delta \zeta^m  =  0\ , \nn
\eea
where $f^\cI_{\cJ \cK} $ are the structure constants of $G$.
For this work it is crucial to realize that there is whole class of higher-dimensional gauge
transformations, with gauge parameters $\hat \Lambda^\cI(x,y)$
and $\hat \Lambda^m(x,y)$  depending non-trivially on $y$, 
which are not included in \eqref{lower-dim-gauge}. We will discuss these 
additional transformations in \autoref{sec:lgts_anomalies} in more detail.

The Coulomb branch of the compactified theory is parametrized by the background values of the scalars $\zeta^\cI$
and $\zeta^m$, by setting 
\beq \label{Coulomb-background}
  \langle \zeta^I \rangle \neq 0 \ , \qquad \langle \zeta^{\boldsymbol \alpha} \rangle = 0 \ , \qquad  \langle \zeta^m \rangle \neq 0 \ ,
\eeq
\textit{i.e.}~giving the Cartan Wilson line scalars a vacuum expectation value. 
This induces the breaking 
\beq
   G \times U(1)^{n_{U(1)}} \rightarrow U(1)^{\rk G} \times U(1)^{n_{U(1)}} \ ,
\eeq
and assigns a mass to the W-bosons $A^{\boldsymbol \alpha}$. Note that 
one has to include in addition the Kaluza-Klein vector $A^0$, such
that the full three- or five-dimensional massless gauge group is actually $U(1)^{\rk G+n_{U(1)} + 1}$.

The massive fields in the lower-dimensional theory are then the excited Kaluza-Klein modes of 
all higher-dimensional states and the fields that acquire masses on the Coulomb branch.
In particular, also the modes of the higher-dimensional 
charged matter states will gain a mass. 
Let us, for example, consider a fermion $\hat \psi$ in the higher-dimensional theory that 
transforms in some representation $R$ under $G$ and has $U(1)$-charges $q_m$ under 
the  gauge fields $\hat{A}^m$. 
This implies that the higher-dimensional covariant derivative takes the form 
\beq \label{def-Dpsi}
  \cD_\mu \hat\psi = \big(\nabla_\mu \hat\psi - i \hat A^\cI_\mu  T^R_\cI   - i q_m \hat A^m_\mu \big) \hat\psi  \, ,
\eeq
We expand $\hat \psi$ in an eigenbasis $\hat \psi(w)$ associated to the weights $w$ of $R$ 
with the property $T^R_I \ \hat\psi(w) = w_I \ \hat \psi(w)$
for the Cartan directions, where
$w_I := \langle \boldsymbol{\alpha}^\vee_I , w \rangle $ are the Dynkin labels  and
$\boldsymbol{\alpha}^\vee_I$ is the simple coroot associated to $T_I$. We refer to \autoref{app:Lieconventions}
for our conventions for Lie groups.
Dimensionally reducing such a fermion $\hat \psi(w,q)$ on a circle, one can infer its 
Coulomb branch mass  $m_{\rm CB}^{w,q}$ in the background \eqref{Coulomb-background}.
Using \eqref{def-Dpsi} we read off 
\beq
   m_{\rm CB}^{w,q} = w_I  \langle \zeta^I \rangle + q_m  \langle \zeta^m \rangle\ ,
\eeq 
In total the mass of a field $\psi_{(n)}(w,q)$ at Kaluza-Klein level $n$ in the
lower-dimensional theory reads
\beq \label{e:KK-masses}
 m = m^{w,q}_{\rm CB} + n \,m_{\rm KK} = w_I   \langle\zeta^I \rangle + q_m  \langle\zeta^m \rangle + \frac{n}{\langle r\rangle},
\eeq
 with $m_{\rm KK} = 1/\langle r \rangle$ being the unit Kaluza-Klein 
mass determined by the background value of the radius. 
Note that a similar analysis can be performed for the Kaluza-Klein modes 
of all other fields, including scalars, W-bosons, and six-dimensional tensor fields.

In particular, let us denote by $T_{\rm sd}$ and 
$T_{\rm asd}$ the number of six-dimensional self-dual and 
anti-self-dual tensors respectively. 
Each such tensor $\hat B^\alpha$, $\alpha = 1,\ldots, T_{\rm sd}+T_{\rm asd}$,
yields a whole tower of Kaluza-Klein states after compactification on the circle. 
While the massive modes are genuine tensor fields in five dimensions \cite{Townsend:1983xs,Bonetti:2012fn},
the massless mode can be dualized into a massless five-dimensional vector field $A^\alpha$.\footnote{
Note that the index $\alpha$ counts the number of massless tensor fields in 
six dimensions and should not be confused with $\boldsymbol{\alpha}$ labelling the roots of $G$.} 
To be more precise, the Kaluza-Klein ansatz for $\hat B^\alpha$ reads  
\beq \label{hatBalpha_red}
  \hat B^\alpha = B^\alpha - \big( A^\alpha - 2 \lambda^{-1}_G b^\alpha \text{tr}_{\rm f}(\zeta A) - 2 b^\alpha_{mn} \zeta^m A^n \big) \wedge Dy\ .
\eeq
The group-specific coefficient is given by 
$\lambda^{-1}_G= \langle {\boldsymbol \alpha}_{\rm max}, {\boldsymbol \alpha}_{\rm max} \rangle / 2$, where $ {\boldsymbol \alpha}_{\rm max}$ is the root of maximal length (see \autoref{app:Lieconventions}). 
The trace $ \text{tr}_{\rm f}$ is evaluated in the fundamental representation of $G$. 
Note that the modification of this ansatz with terms proportional to the constant 
Green-Schwarz coefficients $b^\alpha$ and $b^\alpha_{mn}$, to be defined below in \eqref{GS6}, is important since 
the six-dimensional tensors have modified field strengths (see \textit{e.g.}~\cite{Bonetti:2011mw,Grimm:2013oga} for a more complete discussion). 
They therefore can participate non-trivially in the Green-Schwarz mechanism as we discuss momentarily.
In the classical five-dimensional Coulomb branch parametrized by \eqref{Coulomb-background} the ansatz 
\eqref{hatBalpha_red} including only the massless fields becomes
\beq \label{hatBalpha_red_CB}
   \hat B^\alpha = B^\alpha - \big( A^\alpha - 2   b^\alpha \cC_{IJ} \zeta^I A^J - 2 b^\alpha_{mn} \zeta^m A^n \big) \wedge Dy\ ,
\eeq
where we have introduced the coroot intersection matrix 
$\cC_{IJ} = \lambda^{-1}_G \text{tr}_{\rm f}(T_I T_J)$ with $T_I$ being 
the Cartan generators in the coroot basis. 
Again, we refer to \autoref{app:Lieconventions}, in particular \eqref{e:def_coroot_int_mat} 
and the following text for more details.
In five dimensions the $B^\alpha$ can then be eliminated from the action in favour of 
$A^\alpha$ by using the self- or anti-self-duality of $\hat B^\alpha$.

It will be important in the following that a classical 
four-dimensional or six-dimensional gauge theory does not 
necessarily have to be gauge invariant. In fact, it is well-known 
that often classical gauge non-invariance is required to 
cancel one-loop gauge anomalies induced by chiral fields. Famously, this 
is done by the Green-Schwarz mechanism \cite{Green:1984sg,Sagnotti:1992qw,Sadov:1996zm}. 
In six dimensions the relevant terms are given by\footnote{In general there exist also gravitational Green-Schwarz terms in order to cancel
gravitational anomalies. However, these are of no importance
in our discussion, and we omit them in the following.}
\beq \label{GS6}
    \hat S^{(6)}_{\rm GS} = - \int  \eta_{\alpha \beta } \hat B^\beta\wedge 
    \Big(b^\alpha \lambda^{-1}_G \text{tr}_{\rm f} (\hat F \wedge \hat F)  + b^\alpha_{mn}  \hat F^m \wedge \hat F^n  \Big)\ ,
\eeq
where $b^\alpha$ and $b^\alpha_{mn}$ are the Green-Schwarz anomaly coefficients. 
The matrix $\eta_{\alpha \beta}$ is constant, symmetric in its indices and its signature consists of $T_{\rm sd}$
positive signs and $T_{\rm asd}$ negative ones. 
Anomaly cancellation takes place if the tree-level diagram involving the tensors $\hat B^\alpha$ 
cancels the non-vanishing contribution of all anomalous one-loop diagrams induced by chiral fields 
of the theory. In six dimensions chiral fields are spin-1/2 and spin-3/2 fermions as well 
as self-dual or anti-self-dual tensors.

In four space-time dimensions the Green-Schwarz mechanism can be mediated by 
axions $\hat \rho_\alpha$, $\alpha = 1,\ldots ,n_{\rm ax}$, rather than 
tensor fields. The relevant classical counter-term is then given by
\beq \label{GS4}
   \hat S^{(4)}_{\rm GS} =  - \frac{1}{4}\int   \tensor{\eta}{_\alpha^\beta}\hat\rho_\beta \Big( b^\alpha \lambda^{-1}_G \text{tr}_{\rm f} (\hat F \wedge \hat F)  + b^\alpha_{mn}  \hat F^m \wedge \hat F^n  \Big)\ ,
\eeq
where $ \tensor{\eta}{_\alpha^\beta}$ is a constant square matrix.
This term is non-gauge-invariant if the axions $\hat \rho_\alpha$ are shift-gauged by 
the $U(1)$ gauge fields $\hat A^m$. In analogy to six dimensions anomaly cancellation 
takes place if the tree-level diagram involving the axions $\hat \rho_\alpha$ cancels
the non-vanishing contribution of all anomalous one-loop diagrams induced by chiral fields 
of the theory. In four dimensions such chiral fields are spin-1/2 fermions. 
It is important to point out that after compactification to three space-time dimensions 
the Kaluza-Klein zero-modes of the axions $\hat \rho_\alpha$ can be dualized 
into three-dimensional vectors $A^\alpha$ (see \textit{e.g.}~\cite{Grimm:2010ks,Cvetic:2012xn} for a detailed discussion).

We thus realize that both the six-dimensional and four-dimensional setting 
is characterized by Green-Schwarz coefficients $b^\alpha, b^\alpha_{mn}$. 
Furthermore, the fields $\hat B^\alpha$ and $\hat \rho_\alpha$ appearing 
in \eqref{GS6} and \eqref{GS4} are both captured by vectors $A^\alpha$ in the five- and three-dimensional 
effective theories respectively. This slight abuse of notation will allow us to 
consider both the six-dimensional and four-dimensional case simultaneously. 
Let us also stress that all following considerations 
also apply to situations where no Green-Schwarz mechanism is employed to
cancel anomalies.

\subsection{Large gauge transformations on the Coulomb branch}\label{sec:lgts_anomalies}

In this subsection we discuss in detail the set of gauge transformations of 
an Abelian or non-Abelian theory on a circle that are translated to a symmetry 
of the geometry of an elliptic fibration. Recall that after compactification 
the effective theory admits \eqref{lower-dim-gauge} as
local symmetries before pushed to the Coulomb branch. In the Coulomb 
branch one simply has a purely Abelian local symmetry.

In addition to the lower-dimensional gauge transformations \eqref{lower-dim-gauge}
we could also have performed a circle-dependent gauge transformation 
and then compactified on the circle $y \sim y+2\pi$. If one preserves the boundary conditions of 
the fields in the compactification ansatz the gauge invariance of the higher-dimensional theory then 
implies that there exists a variety of equivalent  
lower-dimensional effective theories that are obtained after circle reduction of 
the same higher-dimensional theory. 
Gauge transformations that cannot 
be deformed continuously to the identity map are known as \textit{large} gauge transformations.
More concretely, let us consider the effect of a gauge transformation that locally takes the form
\beq \label{LambdaAnsatz}
    \hat \Lambda^\cI(x,y) = \left\{\begin{array}{c}
                             - k^I y\\
                             0
                            \end{array} \right. \ , \qquad \hat \Lambda^m(x,y) = - k^m y\ ,
\eeq
where $k^I$ and $k^m$ are constants and we have included a minus sign for later convenience.  $k^I,k^m$ will be further restricted below to ensure that \eqref{LambdaAnsatz} is 
in fact a large gauge transformation (preserving the boundary conditions of all fields).
Using the split $\cI = (I ,\boldsymbol{\alpha} )$ as in \eqref{hatA_expand} 
we have set $\hat \Lambda^{\boldsymbol{\alpha}}(x,y)=0$ to ensure that the Coulomb branch values $ \langle \zeta^{\boldsymbol \alpha} \rangle = 0$ in  \eqref{Coulomb-background} are unchanged. This guarantees that we stay on the 
considered Coulomb branch; all the following discussions are performed on this background. 
The reduction 
ans\"atze \eqref{red-AI} and  \eqref{hatBalpha_red_CB} are also compatible with a gauge transformation \eqref{LambdaAnsatz} if one introduces 
the new quantities 
\beq \label{eq:LGTgeneral_1}
   \tilde r = r\ , \qquad \tilde \zeta^I = \zeta^I + \frac{k^I}{r} \ ,  \qquad \tilde \zeta^m = \zeta^m + \frac{k^m}{r} \ , 
\eeq
and 
\begin{align}
\label{e:non_abelian_trafo_vectors_4d}
 \begin{pmatrix}
  \tilde A^{0}\\[8pt]
  \tilde A^{I}\\[8pt]  
  \tilde A^{m}\\[8pt]
  \tilde A^{\alpha}
 \end{pmatrix} = 
\begin{pmatrix*}[c]
 1 & 0 & 0  &0 \\[8pt]
 - k^I & \delta^{I}_{J} & 0 & 0\\[8pt]
  - k^m & 0 & \delta_n^m & 0 \\[8pt]
 \frac{1}{2}k^K k^L \cC_{KL} b^\alpha +\frac{1}{2}k^p k^q b^\alpha_{pq} & 
 -k^{K} \cC_{KJ} b^\alpha
  &- k^p b^\alpha_{pn}& \delta_\beta^\alpha
\end{pmatrix*}\cdot
\begin{pmatrix}
  A^{0} \\[8pt]
  A^{J} \\[8pt]
  A^{n} \\[8pt]
  A^{\beta} \, 
 \end{pmatrix}.
\end{align}
With \eqref{LambdaAnsatz} being compatible with \eqref{red-AI} and  \eqref{hatBalpha_red_CB}
we mean that the form of the reduction ansatz after a gauge transformation 
is unchanged when using the tilded quantities. 

Some additional remarks are in order here. First, it is important to stress that the simple shifts in the vector fields 
$\tilde A^I$ only occur for the Cartan direction. 
In the non-Cartan directions, \textit{i.e.}~for the vectors that are massive in the Coulomb 
branch, the non-Abelian structure of $G$ modifies the transformation rule. Second, while 
$ \langle \zeta^{\boldsymbol \alpha} \rangle = 0$ is preserved by \eqref{LambdaAnsatz} the actual 
values for $\langle \zeta^{I} \rangle$ do change in the vacuum. One therefore relates theories 
at different points
on the Coulomb branch. In fact, this is a defining property of a large gauge transformation: they
relate theories at different points in the vacuum manifold of the theory with the same properties, see  \textit{e.g.}~\cite{Harvey:1996ur}.  Third, later on we will consider six-dimensional theories with $(1,0)$ supersymmetry
arising in F-theory. These theories have $T_{\rm sd}=1$ and $T_{\rm asd}\equiv T$. 
Each six-dimensional anti-self-dual tensor is accompanied by a real 
scalar field in the multiplet. After dimensional reduction these scalar fields combine with $T$ of the $A^\alpha$ into 
vector multiplets. Importantly, it was found in \cite{Bonetti:2011mw,Grimm:2013oga} that the redefinition of the five-dimensional scalar fields 
is precisely of the form compatible with \eqref{e:non_abelian_trafo_vectors_4d} (see \textit{e.g.}~(3.30) in \cite{Grimm:2013oga}). In other words, a gauge transformation \eqref{LambdaAnsatz} shifts both the vectors and scalars in a compatible fashion. 
We note that a similar story applies to circle compactifications from four to three space-time dimensions. 
In fact, the transformations \eqref{eq:LGTgeneral_1} and \eqref{e:non_abelian_trafo_vectors_4d} 
are equally valid for this latter case. As noted above the 
vectors $A^\alpha$ are the duals of the four-dimensional scalars $\hat \rho_\alpha$
appearing in \eqref{GS4}. 

Clearly, a gauge transformation \eqref{LambdaAnsatz} also 
requires to transform the Kaluza-Klein modes of all higher-dimensional charged fields. 
Given a matter state $\psi_{(n)}$ at Kaluza-Klein level $n$ in the representation $R$ of $G$
and with charge $q_m$ under $\hat A^m$ we first proceed as described 
after \eqref{def-Dpsi} and introduce eigenstates $\psi_{(n)}(w,q)$, where $w$ are the weights of $R$.
The transformation \eqref{e:non_abelian_trafo_vectors_4d} mixes these states as
 \begin{align}
\label{e:non_abelian_trafo_charges_4d}
\psi_{(n)}(w,q) \mapsto \psi_{(\tilde n)}(\tilde w,\tilde q) \ ,\qquad \begin{pmatrix}
  \tilde n\\[8pt]
  \tilde w_{I}\\[8pt] 
  \tilde q_{m}\\
 \end{pmatrix} = 
\begin{pmatrix*}[c]
 1 & - k^J  & - k^n \\[8pt]
 0 & \delta_I^J &0  \\[8pt] 
 0 & 0 & \delta_m^n \\[8pt]
\end{pmatrix*}\cdot
\begin{pmatrix}
  n\\[8pt]
  w_{J}\\[8pt]
  q_n
 \end{pmatrix} \, .
\end{align}
Note that in general this transformation shifts the whole Kaluza-Klein
tower, but there is still no state charged under $\tilde A^{\alpha}$.
Furthermore, imposing that \eqref{e:non_abelian_trafo_charges_4d}
is in fact a consistent reshuffeling of the Kaluza-Klein states, which is 
necessary for invariance of the theory, imposes 
conditions on the constants $k^I$ and $k^m$ that are dependent 
on the spectrum of the theory. 
We will discuss the various choices and conditions 
in the following. 
\vspace{.2cm}

\noindent
\textbf{Integer large gauge transformations}

\noindent
In \eqref{LambdaAnsatz} we have introduced gauge transformations 
that depend on the circle coordinate $y \sim y + 2\pi $. As mentioned before, these 
correspond to  large gauge transformations around the circle if they preserve the 
circle boundary conditions of all fields and wind at least once around the circle. 
Let us now define what we mean by integer large gauge transformations. 
First, if we consider pure gauge theory without charged matter, 
we call a large gauge transformation to be integer if all 
$k^I$ and all $k^m$ are integers. 
Indeed, the degrees of freedom in \eqref{LambdaAnsatz} are in general characterized by
elements in the homotopy groups
 \beq
    \pi_1(U(1)^{{\rm rk} G}) \cong \mathbb{Z}^{{\rm rk} G}\ , \qquad   
    \pi_1(U(1)^{n_{U(1)}}) \cong \mathbb{Z}^{n_{U(1)}}\,.
 \eeq
Clearly, \eqref{LambdaAnsatz} define maps from $S^1$ into the gauge group (which is purely 
Abelian on the Coulomb branch). These are precisely classified by the first homotopy group of 
the gauge group, which in the case at hand consists of tuples of integers.

If one now includes a charged matter spectrum the invariance of the 
boundary conditions of all these fields dictates the set of large gauge 
transformations. In these cases, the space of allowed 
$k^I$ and $k^m$ has to be quantized. In general, the $k^I$ 
and $k^m$ could still be integer or fractional depending on the 
weights and charges of the matter fields. 
However, for the transformations \eqref{LambdaAnsatz} to be an actual symmetry, 
{i.e.}~a large 
gauge transformation, the following condition for each state $\hat \psi(R,q)$ has to be satisfied:
\beq \label{kk_cond}
  k^I w_I + k^m q_m \quad \in \ \mathbb{Z}\ , 
\eeq
where $w_I$ are the weights of $R$ and $q_m$ are the $U(1)$ charges.
This condition also arises from the transformation of the Kaluza-Klein level 
in \eqref{e:non_abelian_trafo_charges_4d} and ensures that $\tilde n$ is an integer, which 
implies equivalence of the full Kaluza-Klein towers of the compactified theory by a simple reshuffling.
Now we are in the position to introduce our notion of integer large gauge transformations. 
They are spanned by pairs  $(k^I , k^m)$ satisfying \eqref{kk_cond} and one of the conditions  
\begin{itemize}
 \item[(I)] $k^m=0$ and $k^I \in \bbZ$,
 \item[(II)] $k^m \in \bbZ$ and  $k^I w_I \in \mathbb{Q}$.
\end{itemize}
It is useful to comment on the class (II) of basis vectors. 
While all $w_I$ reside in an integer lattice 
and therefore do not violate \eqref{kk_cond} for integer $k^I$,
the $U(1)$ charges $q_m$ can be fractional. However, we will also consider the set of integer $k^m$'s that 
allow a compensation of this fractional contribution to \eqref{kk_cond} by an appropriate 
$k^I$-transformation which might be fractional. 
\vspace{.2cm}

\noindent
\textbf{Special fractional large gauge transformations}

\noindent
There is another set of large gauge transformations that will be of importance for us. 
If the arising representations in the spectrum of matter states is special, {e.g.}~if the fundamental representation does not occur, also 
fractional $k^I$ might be allowed. 
More precisely, we also want to consider pairs $(k^I , k^m)$ satisfying \eqref{kk_cond}
and
\begin{itemize}
 \item[(III)] $k^m=0$ and $k^I$ fractional. 
\end{itemize}
We call large gauge transformations satisfying (III) special fractional large gauge transformations.
Note that the conceptual difference between (III) and (I), (II) is that there is always at least one integer quantity $k^m, k^I$ in (I) and (II).

It remains to consider the cases where also $k^m$ is fractional. For example, this could 
be allowed if the spectrum has special charges such that $k^m q_m$ is integer for each state.
We find, however, that such a situation does not occur in our geometric considerations. In the known 
F-theory examples there are always states that have minimal charge $0<q_m \leq 1$.
Following some folk theorems (see {e.g.}~\cite{Banks:2010zn,Hellerman:2010fv}) this might be true in any theory 
of quantum gravity. In this case the space of all large gauge transformations is spanned 
by $(k^I,k^m)$ satisfying \eqref{kk_cond} and either (I), (II) or (III).

\newpage

\noindent
\textbf{Comments on one-loop Chern-Simons terms}

\noindent
The listed large gauge transformations are of key importance if 
one aims to study the anomalies of the six- or four-dimensional 
theory by using the circle compactified effective theory and its
Chern-Simons terms. Let us combine all massless vectors in the Coulomb branch as 
$A^\Sigma= (A^0,A^I,A^m,A^\alpha)$. In five dimensions the gauge Chern-Simons terms
then take the form
\beq
  S^{(5)}_{CS} = -\frac{1}{12}\int k_{\Sigma \Lambda \Gamma} A^\Sigma \wedge F^\Lambda \wedge F^\Gamma\ , 
\eeq
while in three dimensions they are given by
\beq
   S^{(3)}_{CS} = \int \Theta_{\Sigma \Lambda} A^\Sigma \wedge F^\Lambda
\eeq
The coefficients $k_{\Sigma \Lambda \Gamma} $ and $\Theta_{\Sigma \Lambda}$
are constants and can either arise directly in the dimensional reduction or be induced 
at one loop after integrating out all massive states. It was a key result of \cite{Grimm:2015zea}
that the one loop Chern-Simons terms precisely yield the higher-dimensional
gauge anomaly conditions when transformed under \eqref{e:non_abelian_trafo_vectors_4d}
and \eqref{e:non_abelian_trafo_charges_4d} (see \cite{Cvetic:2012xn,Grimm:2013oga}
for earlier results on the four- and six-dimensional (mixed) gravitational anomalies) . 
This can be traced back to the fact that 
the large gauge transformations are only a symmetry at  one-loop if anomalies are cancelled. 
It appears that the Chern-Simons terms of the lower-dimensional effective 
theory capture all the required data to perform this test for the underlying higher-dimensional 
theory.

\section{On the systematics of F-theory compactifications} \label{Ftheoryfacts}

In the previous sections we have discussed the compactification of 
gauge theories on a circle. Such theories prominently arise in the context of F-theory compactifications.
In order to derive the effective action of F-theory on an elliptically-fibered Calabi-Yau geometry 
one is forced to use the duality to M-theory. 
More precisely, F-theory on a singular elliptically-fibered Calabi-Yau manifold,
after an additional circle-reduction and moving
to the Coulomb branch is dual to M-theory on the same but resolved manifold. Shrinking the elliptic fiber to zero size on the
M-theory side corresponds to the decompactification limit of the circle on the F-theory side.
Many properties of the F-theory effective action, like gauge symmetry, parts of the spectrum and Yukawa couplings,
are encoded in the geometry of the elliptic fibration. A profound understanding of the latter therefore offers deep insights into the heart of F-theory compactifications. Noting that the M-theory to F-theory limit provides the correct approach 
to understanding the system, we have to suspect that the smooth Calabi-Yau geometry should share the 
symmetries of the gauge theories on a circle introduced in \autoref{sec:lgts_anomalies}. 
In the following we will therefore introduce some 
basics about the smooth F-theory geometries relevant to this work. 

For the rest of this work we will study six-dimensional theories with $\cN=(1,0)$ and four-dimensional theories
with $\cN=1$ supersymmetry. These arise as effective actions of F-theory on elliptically fibered Calabi-Yau
threefolds and fourfolds, respectively. Non-Abelian gauge symmetries are induced from singularities of the 
elliptic fiber at codimension-one loci in the base, while Abelian gauge factors are related to rational sections.
The effective actions are accessed using the M-theory to F-theory limit. The circle reduced theories are thus
compared with M-theory, or rather eleven-dimensional supergravity, compactified on the smooth Calabi-Yau manifold \cite{Vafa:1996xn,Morrison:1996na,Morrison:1996pp,Ferrara:1996wv,Denef:2008wq,Grimm:2010ks,Bonetti:2011mw,Grimm:2013oga}. 
In this manner
the four-/six-dimensional supergravity data of the F-theory effective action is matched to
geometric quantities using the three- or five-dimensional effective theories, respectively. 
It is important to realize that a classical matching procedure fails. In particular, for certain Chern-Simons terms on the M-theory side one cannot find counterparts in the circle reduced supergravity at the classical level.
The required corrections in the circle compactification arise at the one-loop level after integrating 
out the massive modes. As discussed briefly at the end of \autoref{sec:lgts_anomalies} this process induces 
one-loop Chern-Simons terms and both reductions can be matched properly \cite{Grimm:2011fx,Bonetti:2011mw,Cvetic:2012xn,Grimm:2013oga,Cvetic:2013uta,
Anderson:2014yva}.
We highlight that, since one-loop induced Chern-Simons terms carry information about the number of 
matter fields, the matching to M-theory allows to translate information about the spectrum into the geometric 
data of the resolved space. This geometric perspective will be introduced next. 

First let us establish some geometric notions of the Calabi-Yau manifold. 
We denote the resolved Calabi-Yau space by $\hat Y$. We assume that it constitutes an elliptic fibration over some base space
$B$, with the corresponding projection given by $\pi: \hat Y \rightarrow B$. For this subsection we will assume that 
the fibration has at least one section. 
A set of linearly independent (minimal) rational sections of the elliptic fibration is 
denoted by $s_0$, $s_m$, $t_r$ where one arbitrary section $s_0$
is singled out as the so-called zero-section. The sections $t_r$ will be purely torsional, while the $s_m$
are assumed to be non-torsional. We will have to say more about rational sections and 
this distinction in \autoref{sec:ec_group_structures}.
Furthermore, there might exist a divisor $S^{\rm b}$ in the base $B$ of the resolved space $\hat Y$ 
over which the fiber becomes reducible with the individual irreducible components
intersecting as the (affine) Dynkin diagram of the gauge algebra. Fibering these
over the corresponding codimension-one locus $S^{\rm b}$ in $B$ yields the blow-up divisors of $\hat Y$, 
which we denote by $D_I$.

In the following we define a basis of divisors $D_\Lambda = (D_0 , D_I , D_m , D_\alpha)$ on the resolved space $\hat Y$ 
in the correct frame
such that the
corresponding gauge fields obtained from the expansion of the M-theory three-form
\begin{align}\label{e:expansion_threeform}
 C_3 = A^0\wedge [D_0] + A^m\wedge [D_m]+ A^I\wedge [D_I] + A^\alpha\wedge [D_\alpha]
\end{align}
can be matched properly to the circle reduced theory. In this expression $[D]$ denotes the Poincar\'e-dual 
two-form to the divisor $D$ in $\hat Y$.

\begin{itemize}
 \item Divisors $D^{\rm b}_\alpha$ 
 of the base $B$ define the vertical divisors $D_\alpha:=\pi^{-1}(D^{\rm b}_\alpha)$ via pullback.
 For each $D^{\rm b}_\alpha$ in $B$ there is an axion in the four-dimensional F-theory compactification and an (anti-)self-dual tensor in the
 six-dimensional setting, respectively. Supersymmetry implies $T_{\rm sd} = 1$ and thus we find 
 \bea
     T_{\rm asd} &\equiv & T = h^{1,1}(B)-1\ \ \textrm{in six dimensions}\ , \\
     n_{\rm ax} &=& h^{1,1}(B)\ \ \textrm{in four dimensions}\ , \nn
 \eea
 with $T_{\rm sd}, T_{\rm asd}, n_{\rm ax}$ as defined in \autoref{sec:gensetup+reduction}.

 For Calabi-Yau fourfolds it is also necessary to introduce vertical four-cycles $\cC^\alpha:=\pi^{-1}(\cC_{\rm b}^\alpha)$, which
are the pullbacks of curves $\cC_{\rm b}^\alpha$ in the base intersecting the $D^{\rm b}_\alpha$ as
\begin{align}\label{e:def_metric}
 \tensor{\eta}{_\alpha^\beta} = D^{\rm b}_\alpha \cdot \cC^\beta_{\rm b}
\end{align}
with $\tensor{\eta}{_\alpha^\beta}$ a full-rank matrix. 
For Calabi-Yau threefolds the analogous intersection matrix
\begin{align} \label{etaalphabeta}
 \eta_{\alpha\beta} := D^{\rm b}_\alpha \cdot D^{\rm b}_\beta
\end{align}
is used to raise and lower indices $\alpha, \beta$.
The matrices \eqref{e:def_metric} and \eqref{etaalphabeta} appear in the four- and six-dimensional Green-Schwarz terms \eqref{GS4} and \eqref{GS6}, respectively. 
 
 For later convenience we also define the projection of two arbitrary divisors $D, D^\prime$ as
\begin{align}
 \pi (D\cdot D^\prime) :=
 \begin{cases}
 \big( D \cdot D^\prime \cdot \cC^\beta\big)\ \tensor{\eta}{^{-1}_\beta^\alpha}\ D_\alpha   &\textrm{in three dimensions,} \\
 \big(D \cdot D^\prime  \cdot D^\alpha\big) D_\alpha  & \textrm{in five dimensions.}
 \end{cases}
\end{align}
Furthermore we denote by $\pi_{\cM I}$ the intersection number of a section $s_{\cM}$ with a blow-up divisor $D_I$ in the elliptic fiber $\mathcal E$:
\begin{align}
 \pi_{\cM I} := \cap (s_{\cM}, D_I)\big\vert_{\mathcal E}\, .
\end{align}  

 \item We denote the divisor associated to the zero-section $s_0$ by $S_0\equiv Div(s_0)$. 
 The divisor $D_0$ is then defined by shifting $S_0$ as
 \begin{align}\label{e:old_base}
 D_0 = S_0 - \frac{1}{2}\pi(S_0 \cdot S_0)\ .
\end{align}
 The corresponding vector $A^0$ in \eqref{e:expansion_threeform} is identified with the Kaluza-Klein vector in the 
 circle reduced F-theory setting, {i.e.}~with $A^0$ in \eqref{e:metric_reduction}. 
 
 \item The $D_I$ denote the blow-up divisors and yield the 
 Cartan gauge fields $A^I$ in \eqref{e:expansion_threeform}. This implies that $I=1,\dots , \rk G$.

 \item Given a set of rational sections $s_m$ 
 the $U(1)$ divisors $D_m$ are defined via the Shioda map. Denote by $S_m \equiv {Div}(s_m)$ the 
 divisor associated to $s_m$. The Shioda map ${D}(\cdot)$ reads
 \beq\label{e:old_shioda}
 D(s_m) \equiv D_m = S_m - S_0 - \pi \Big (  (S_m -
 S_0  ) \cdot S_0 \Big )
 + \pi_{mI} \, \mathcal{C}^{-1\,IJ}D_J  \, ,
\eeq
where $\cC_{IJ}$ is the coroot intersection matrix \eqref{e:def_coroot_int_mat} of $G$.
 The $D_m$ yield in \eqref{e:expansion_threeform} the 
 Abelian gauge symmetries in the F-theory setting such that $m=1,\dots , n_{U(1)}$.
 
 \item The crucial property of the purely torsional sections $t_r$ is that they have 
 no non-trivial image under the Shioda map. Denoting the divisors associated to $t_r$ 
 by $T_r = Div(t_r)$ one has \cite{Mayrhofer:2014opa}
 \beq \label{Shioda_tr}
  D(t_r) = T_r - S_0 - \pi \Big (  (T_r -
 S_0  ) \cdot S_0 \Big )
 + \pi_{mI} \, \mathcal{C}^{-1\,IJ}D_J = 0\ ,
 \eeq
 which, as the other expressions above, should be read in homology.
\end{itemize}
By the Shioda-Tate-Wazir theorem $(D_0 , D_I , D_m , D_\alpha)$ indeed form a basis of the Ner\'on-Severi group of divisors\footnote{For
Calabi-Yau manifolds
the Ner\'on-Severi group coincides with the Picard group, which is why we will identify both groups throughout this paper.} (times $\mathbb Q$).
Note also that the definition of the base divisor ensures
\begin{align}\label{e:base_orth}
 \pi(D_0 \cdot D_0) = 0
\end{align}
and the Shioda map enjoys the orthogonality properties
\begin{align}\label{e:shioda_orth}
 \pi(D_m \cdot D_\alpha) = \pi(D_m \cdot D_I) = \pi(D_m \cdot D_0) = 0 \, ,
\end{align}
which are essential in order to perform the F-theory limit correctly. The blow-up divisors 
further satisfy the properties 
\beq \label{blowup-cond}
   \pi (D_I \cdot D_\alpha) = \pi(D_I \cdot D_0) = 0\ . 
\eeq

Via the matching of the M-theory compactification to the circle reduced theory the intersections of the divisor basis
$\pi (D_\Lambda \cdot D_\Sigma) \equiv  \pi (D_\Lambda \cdot D_\Sigma)^\alpha D_\alpha$ can be nicely 
related to four- and six-dimensional supergravity data
\begin{subequations}\label{evaluate_DD}
\begin{align} 
 -\pi(D_I \cdot D_J)^\alpha &=
   \cC_{IJ}\ b^\alpha \, , \\
 -\pi(D_m \cdot D_n)^\alpha &=
    b^\alpha_{mn} \, ,
\end{align}
\end{subequations}
where  $b^\alpha, b^\alpha_{mn}$ are the Green-Schwarz couplings appearing in \eqref{GS4} and \eqref{GS6}.
These relations hold both for Calabi-Yau three- and fourfolds. 
The Green-Schwarz coefficients $b^\alpha$ are equivalently obtained 
as 
\beq
     S^{\rm b} = b^\alpha D_\alpha^{\rm b}\ , 
\eeq
where $S^{\rm b}$ was the divisor in $B$ supporting the non-Abelian gauge group.  
The remaining intersections are not directly relevant in the following discussions and may be found,
for example, in \cite{Grimm:2013oga,Cvetic:2013uta,Grimm:2015zea}.

\section{Arithmetic structure on fibrations with rational sections} \label{sec:ec_group_structures}

In this section we argue that the arithmetic structures of 
elliptic fibrations with multiple rational sections correspond to certain 
large gauge transformations introduced in \autoref{sec:lgts_anomalies}.  
The considered arithmetic is encoded by the so-called \textit{Mordell-Weil 
group of rational sections}, which we introduce in more detail in \autoref{MWgeneralities}.
In \autoref{MWgeneralities} we also discuss how the geometric 
Mordell-Weil group law translates to a general group law for rational sections 
in terms of homological cycles.
The free generators of Mordell-Weil group correspond 
to Abelian gauge symmetries in the effective F-theory action. 
In \autoref{sec:FreeMWShifts} we show that group actions of the free part of the Mordell-Weil group 
are in one-to-one correspondence to specific integer large gauge transformations along the F-theory circle.
A similar analysis for the torsion subgroup is performed in \autoref{sec:MWtorsion}. 
We find that it precisely captures special fractional non-Abelian gauge transformations 
introduced in \autoref{sec:lgts_anomalies}, which indicates the presence of a non-simply connected
non-Abelian gauge group.

\subsection{On the Mordell-Weil group and its divisor group law} \label{MWgeneralities}
 
A famous arithmetic structure on an elliptic curve is encoded by the Mordell-Weil group.
The Mordell-Weil group is formed by the rational points of an elliptic curve endowed 
with a certain geometric group law (see, \textit{e.g.}~\cite{Silverman}).
The rational points on the generic elliptic fiber of an elliptic fibration $Y$ directly extend to rational sections
and form a finitely generated Abelian group, which is called the Mordell-Weil group of rational sections
$\textrm{MW}(Y)$. Thus it splits into a free part and a torsion subgroup
\begin{align} \label{splitMW}
 \textrm{MW}(Y) \cong \bbZ^{\rk \textrm{MW}(Y)} \oplus \bbZ_{k_1} \oplus \dots \mathbb 
    \oplus \bbZ_{k_{n_{\rm tor}}}  \, .
\end{align}
Having chosen one (arbitrary) zero section as the neutral element of the Mordell-Weil group, $\rk \textrm{MW}(Y)$ rational sections generate the free part
and $n_{\rm tor}$ rational sections generate the torsion subgroup. The precise group law on the generic fiber (in Weierstrass form) may be looked up for example
in \cite{Silverman}. We denote the addition of sections $s_1,s_2$ 
using the Mordell-Weil group law by `$\oplus$', \textit{i.e.}~we write $s_3= s_1 \oplus s_2$ with $s_3$ being the new
rational section.
Since, as noted before, the rational sections $s_\cM$ of an elliptic fibration define 
divisors $S_\cM \equiv {Div}(s_\cM)$, we will investigate how the group law is translated to divisors.
More precisely, we will derive the divisor class
\begin{align}
 {Div}( s_1 \oplus k s_2 )\, , \quad k \in \mathbb Z \, ,
\end{align}
where $k s_2 = s_2 \oplus \ldots \oplus s_2$ with $k$ summands. 
In 
contrast the addition in homology of divisor classes associated to sections
is denoted by `$+$' .
Extending the treatment in \cite{Morrison:2012ei} the group law, written in homology, is uniquely determined by the three conditions:
\begin{enumerate}
 \item The Shioda map $D(s_{\cM})$ introduced in  \eqref{e:old_shioda} is a homomorphism from the Mordell-Weil group to the Ner\'on-Severi group (times $\mathbb Q$)
 \begin{align}
  D(s_1 \oplus ks_2) = D(s_1) + k\, D(s_2) \, .
 \end{align}

 \item A section $s_\cM$ intersects the generic fiber $\mathcal E$ exactly once
 \begin{align}
  S_\cM \cdot \mathcal E = 1 \, .
 \end{align}
 \item In the base $B$ a divisor $S_\cM$ associated to a section squares to the canonical class of the base $K$, {i.e.}
 \begin{align}
  \pi (S_\cM \cdot S_\cM) = K \, .
 \end{align}
\end{enumerate}
The group law for two sections $s_1, s_2$ on the level of divisors then  takes the form
\begin{align}\label{e:MW_law}
 {Div}(s_1 \oplus k s_2 ) &= S_1 + k(S_2 - S_0)
 - k\pi \Big( ( S_1 - k S_0 )\cdot 
  ( S_2 - S_0  ) \Big )  \, ,  
\end{align}
where $S_0$ denotes the divisor associated to the zero-section $s_0$. We stress that we assumed that blow-up divisors do not contribute to the group-law. Although we could not
derive this from first principles, their inclusion leads to inconsistencies in the effective field theory when considering F-theory compactifications.

It is well known that the Shioda map as an injective homomorphism \eqref{e:old_shioda} transfers this group structure
to the Ner\'on-Severi group (times $\mathbb Q$) of divisors modulo algebraic equivalence.
Therefore it is reasonable to ask how a Mordell-Weil
group action on the elliptic fibration effects the circle-reduced supergravity. We will find that the free part of the Mordell-Weil group
corresponds to certain Abelian large gauge transformations, while the torsion subgroup manifests in special fractional
non-Abelian large gauge transformations.
As briefly mentioned at the end of \autoref{sec:lgts_anomalies} and discussed in more 
detail in \cite{Grimm:2015zea} these arithmetic structures allow to 
establish the cancellation of all pure Abelian and mixed Abelian-non-Abelian gauge anomalies in 
the effective field theory of F-theory.

\subsection{The free part of the Mordell-Weil group}
\label{sec:FreeMWShifts}
 
Let us first consider the free part of the Mordell-Weil group.
On the elements of the Mordell-Weil basis, consisting of the 
zero-section $s_0$, the free generators $s_m$, and the torsional
generators $t_r$, we now perform a number of $k^m \in \mathbb Z$ shifts into the directions of the free generators $s_{m}$,
\textit{i.e.}~we find a new Mordell-Weil basis given by
\begin{align}
\label{eq:MWshift}
 &\tilde s_0 := s_0 \oplus k^n s_{n} \, , &&  \tilde s_m := s_m \oplus k^n s_{n} \, , &&
 \tilde t_r := t_r \oplus k^n s_{n} \, ,
\end{align}
where $ k^n s_{n}=  k^1 s_{1} \oplus \ldots  \oplus k^{n_{U(1)}} s_{n_{U(1)}}$ and 
each summand $k^1 s_1, k^2 s_2, \ldots$ is evaluate using the Mordell-Weil group law.
Our goal will be to translate these shifts to the divisor basis $D_\Lambda = (D_0 , D_I , D_m , D_\alpha)$
introduced in \autoref{Ftheoryfacts}, and then identify the corresponding large 
gauge transformation. 

We use the formula \eqref{e:MW_law} to identify the change in the definition of the $U(1)$ divisors $D_m$, 
the Cartan divisors $D_I$ and the base divisor $D_0$. Note that the divisors $\tilde D_\alpha = D_\alpha$ are unchanged 
under this transformation. Explicitly we find 
\begin{subequations}  \label{e:MW_shift_divisors_1}
\begin{align} 
\label{free_shift_Ab} \tilde D_m &= D_m - k^n \, \pi (D_{n}\cdot D_m)   , \\[.1cm]
\label{DIshift} \tilde D_I &= D_I - k^K \, \pi (D_K\cdot D_I) 
 \ , \\
 \tilde D_0  
& =  D_0 + k^n \,D_{n} + k^J\, D_J   -\frac{k^n k^p}{2} \pi (D_{n}\cdot D_{p}) -\frac{k^J k^L}{2} \pi (D_J\cdot D_L) \ .
\end{align} 
\end{subequations}
where we have defined 
\beq \label{def-associatedkK}
   k^I := - k^n \, \pi_{n J} \,\mathcal{C}^{-1\,JI} \ . 
\eeq
Using the expressions \eqref{evaluate_DD} one further evaluates
\begin{align} \label{D-transform1}
 \begin{pmatrix}
  \tilde D_{0}\\[8pt]
  \tilde D_{I}\\[8pt]  
  \tilde D_{m}\\[8pt]
  \tilde D_{\alpha}
 \end{pmatrix} = 
\begin{pmatrix*}[c]
 1 & k^J & k^n  &  \frac{k^q k^p}{2} b^\beta_{qp}  +\frac{k^J k^L}{2} \cC_{JL} b^\beta \\[8pt]
0 & \delta^{J}_I & 0 &   k^K \cC_{IK} b^\beta \\[8pt]
  0 & 0 & \delta_m^n & k^k b^\beta_{mk} \\[8pt]
 0 & 
 0
  & 0& \delta^\beta_\alpha
\end{pmatrix*}\cdot
\begin{pmatrix}
  D_{0} \\[8pt]
  D_{J} \\[8pt]
  D_{n} \\[8pt]
  D_{\beta} \, 
 \end{pmatrix}.
\end{align}
It is now straightforward to check that inserting \eqref{D-transform1} and the large gauge transformations 
\eqref{e:non_abelian_trafo_vectors_4d} with $(k^m,k^I)$ as in \eqref{eq:MWshift} and \eqref{def-associatedkK} one finds 
\beq
    C_3 = A^\Lambda \wedge [D_{\Lambda}] = \tilde A^\Lambda \wedge [\tilde D_{\Lambda}] \ . 
\eeq
This implies that the Mordell-Weil shift \eqref{eq:MWshift} indeed induces a large gauge transformation 
discussed in \autoref{sec:lgts_anomalies}. More precisely, the 
pair $(k^m,k^I)$ are in general the basis vectors of type (II) and realize an Abelian 
large gauge transformation combined with a fractional non-Abelian large gauge transformation. 
It is also straightforward to check that the quantization condition 
\eqref{kk_cond} is satisfied for the pair $(k^m,k^I)$ inferred from the geometry. 
In fact, one finds 
\beq
   k^I w_I + k^n q_n =  k^n ( - \pi_{n J} \,\mathcal{C}^{-1\,JI} w_I +  q_n)  = k^n \Big(S_n - S_0 - \pi \big (  (S_n -
 S_0  ) \cdot S_0 \big) \Big)\cdot \cC \ ,
\eeq
where we have used that $w_I = D_I \cdot\cC$ and $q_m = D_m\cdot \cC$ are the charges of a matter state $\hat \psi(w_I , q_m)$
arising from a 
the M2-brane wrapped on the curve $\cC$. The condition  \eqref{kk_cond} then follows 
from the fact that $k^m \in \bbZ$ and  
the appearing intersections between divisors $S_n,S_0$ and the curve $\cC$ are always integral. From its definition \eqref{def-associatedkK} it 
is also clear that $k^I$ can be fractional due to the appearance of the inverse $\cC^{-1\, IJ}$.

Let us make a few comments concerning the derivation and 
interpretation of \eqref{e:MW_shift_divisors_1} and \eqref{D-transform1}.
First, it seems from counting the number of conditions \eqref{eq:MWshift} and \eqref{e:MW_shift_divisors_1}
that the former conditions cannot suffice to fix the complete transformation law. In fact, 
the shift of the non-Abelian Cartan divisors $D_I$ to $\tilde D_I$ is not immediately inferred from \eqref{eq:MWshift},
but appears to be crucial to make the transformation well-defined. To derive \eqref{e:MW_shift_divisors_1} one first 
starts with the transformation to $\tilde D_0$ by evaluating $\tilde S_0 = Div(s_0 \oplus k^n s_{n})$ and using 
\eqref{e:old_base}, which is 
straightforward. If one tries to proceed in a similar fashion for $\tilde D_m$ one realizes that the evaluation 
of the Shioda map \eqref{e:old_shioda} for $\tilde D_m$ in the transformed divisors formally 
requires also to use  new $\tilde D_I$, which is not fixed by \eqref{eq:MWshift}. However, note that the 
shift \eqref{DIshift} is uniquely fixed by requiring that the $\tilde D_I$ again behave as genuine blow-up divisors.
More precisely, we find that  \eqref{DIshift} is fixed if 
the three conditions
\begin{align}
 \pi (\tilde D_I \cdot \tilde D_\alpha) &\overset{!}{=} 0 \, , &  \pi (\tilde D_I \cdot \tilde D_J) &\overset{!}{=}  \pi (D_I \cdot D_J) \, , &
 \pi (\tilde D_I \cdot \tilde D_0) &\overset{!}{=} 0 \, .
\end{align}
are to be satisfied for the new divisors. These are simply the conditions \eqref{blowup-cond} and \eqref{evaluate_DD} in the $\tilde D_\Lambda$ basis. Having fixed $\tilde D_I$ the transformed $\tilde D_m$ in \eqref{free_shift_Ab} are determined 
uniquely.

Let us stress again that the non-Abelian part of this gauge transformation is absolutely essential. 
On the one hand it is non-zero if and only if $\pi_{m I} \neq 0$
for some $D_I$. On the other hand we find fractional $U(1)$-charges if and only if $\pi_{m I} \neq 0$. This can easily
be seen in the Shioda map \eqref{e:old_shioda}. The $U(1)$ charge $q_m$ of an M2-brane state wrapping a holomorphic curve $\cC$ is given by
the intersection of $\cC$ with $D_m$. A fractional contribution to the charge therefore can only arise from the last term in \eqref{e:old_shioda}
since $\mathcal{C}^{-1\,IJ}$ in general has fractional components. In fact, since a section can only intersect nodes with Coxeter label
equal to one, the last term
in \eqref{e:old_shioda}
always vanishes for the simple Lie algebras $E_8, F_4, G_2$, which do not have nodes with Coxeter label one, and they are precisely the only simple Lie algebras
having integer $\mathcal{C}^{-1\,IJ}$.\footnote{This is related
to the fact that the center of the corresponding universal covering group is trivial. We will elaborate more on this
fact in the part about the torsion subgroup of the Mordell-Weil group.} 
The Coxeter labels for the simple Lie algebras can be found in \autoref{t:lie_conventions}.
To put it in a nutshell, if and only if there are fractional $U(1)$ charges, the free Mordell-Weil group action induces Abelian large gauge transformations
supplemented by non-zero fractional non-Abelian large gauge transformations. Effectively, in 
the presence of fractional $U(1)$ charges, a pure Abelian large gauge
transformation with integer winding $k^m \in \mathbb Z$ is in general ill-defined. 
What makes it well-behaved is precisely the additional contribution
from the fractional non-Abelian large gauge transformation (which is by itself also ill-defined)
which compensates for the fractional part in the Abelian sector. This matches the gauge theory discussion of \autoref{sec:lgts_anomalies}.

It should be stressed that not all redefinitions \eqref{e:MW_shift_divisors_1} have an immediate geometric interpretation. 
Away from the singular loci in $B$ that are resolved, the transformation  $D_m \rightarrow \tilde D_m$ is induced by the action of the Mordell-Weil group, which has a 
known geometric origin as addition of points in the fiber (see \textit{e.g.}~\cite{Silverman}). We are not 
familiar of how the latter geometric group law is extended to the non-Abelian singularities or to 
their resolutions. This prevents us from identifying a geometric interpretation of $D_I \rightarrow \tilde D_I$. 
Nevertheless the considered arithmetic operations are well-defined on the level of divisors and consistently 
include also the blow-up divisors, for example, in the Shioda map. 
This suffices to infer information about effective theory after compactification on this 
space. We will encounter similar transformations in the general discussion of non-Abelian large gauge 
transformations in \autoref{Arithmetics_nonAb}.

\subsection{The torsion part of the Mordell-Weil group} 
\label{sec:MWtorsion}

In a similar spirit we can show that a non-trivial torsion subgroup in \eqref{splitMW} is connected to special 
fractional non-Abelian large gauge transformations, \textit{i.e.}~to the basis vectors $(k^I,k^m)$ of type (III) introduced
in \autoref{sec:lgts_anomalies}.
For a torsional section $t_r$, $r=1,\ldots,n_{\rm tor}$, we use the key fact that its image under the Shioda map vanishes, {cf.}~\eqref{Shioda_tr}. As discussed in \autoref{Ftheoryfacts} this implies that they do not define divisors that 
appear in the Kaluza-Klein expansion \eqref{e:expansion_threeform} and thus do not give rise to massless gauge fields 
in the effective theory.

 Despite the fact that torsional sections do not define massless gauge fields in the effective field theory, the 
 Mordell-Weil group action along these sections nevertheless results in a non-trivial 
 transformation  in the circle-reduced theory. In order to show this 
 we perform $\tilde k^r \in \mathbb Z$ shifts along the torsional generators $t_r$
 using the Mordell-Weil group law \eqref{e:MW_law}. The derivation 
 proceeds in a similar fashion as the one in \autoref{sec:FreeMWShifts}.
 In fact, keeping in mind that $D(t_r)=0$ one can use \eqref{e:MW_shift_divisors_1} to infer 
\begin{align}
\tilde D_m &= D_m \ ,\\
\tilde D_I &= D_I - k^K  \pi (D_K\cdot D_I) \ , \\ 
 \tilde D_0  
& =  D_0  + k^J D_J   - \frac{k^J k^L}{2} \, \pi (D_J\cdot D_L) \, ,
\end{align}
where we have defined, similar to \eqref{def-associatedkK}, that 
\beq
  k^I := - \tilde k^r \ \pi_{rJ} \,\mathcal{C}^{-1\,JI}\ .
\eeq
Just as in \autoref{sec:FreeMWShifts}, in general $k^I$ will be fractional due to the appearance of the inverse matrix $\mathcal{C}^{-1\,JI}$. 
In other words the transformations induced by $\tilde k^r$ correspond to special fractional 
non-Abelian large gauge transformations parametrized by pairs $(k^I, k^m=0)$, introduced 
as case (III) in \autoref{sec:lgts_anomalies}.

The fact that torsion in the Mordell-Weil group allows for the presence of special fractional 
non-Abelian large gauge transformations is not unexpected. As discussed in \cite{Aspinwall:1998xj,Mayrhofer:2014opa}, torsion in the Mordell-Weil group indicates that the gauge group is 
not simply connected, and therefore certain representations of the Lie algebra do not appear on the level of the group, \textit{i.e.}~the weight 
lattice of the group is coarser. Because of this fact also certain fractional non-Abelian
large gauge transformations are compatible with the circle boundary conditions. Indeed the 
torsional shifts exhaust all possible fractional large gauge
transformations, which is evident from considering the center of the universal covering group as
in section 3.3 of \cite{Mayrhofer:2014opa}.

 \section{Introducing an arithmetic structure on fibrations with multi-sections} \label{sec:multi_group}

In this section we aim to generalize the discussion of \autoref{sec:ec_group_structures} to 
Calabi-Yau geometries that admit a genus-one fibration that does not have a rational section.
These setups always come with multi-sections that no longer cut out rational points of the genus-one 
fiber but rather roots, which are exchanged over branch cuts in the base $B$. On such 
genus-one fibrations with only multi-sections there is no known arithmetic structure analog to
the Mordell-Weil group. However, our understanding of the F-theory effective action 
associated to such geometries, which will have U(1) gauge group factors if there is more than one 
independent multi-section, and the possibility to perform Abelian large gauge transformations in 
these setups suggest that an arithmetic structure should equally exist on genus-one fibrations. 
We will collect evidence for the existence of this structure, which we name the \textit{extended Mordell-Weil group}, and study its key properties.

Our considerations will be driven by two facts. First, we will make use of the fact that 
a genus-one fibration with multi-sections can often be related by a geometric transition to 
elliptic fibrations with multiple rational sections. Physically this corresponds to an unHiggsing 
of Abelian gauge fields \cite{Morrison:2014era,Anderson:2014yva,Klevers:2014bqa,Garcia-Etxebarria:2014qua,Mayrhofer:2014haa,Mayrhofer:2014laa,Cvetic:2015moa}. Following the divisors through this transition we are able 
to reverse-engineer, on the level of divisor classes, the group law on the genus-fibration from the 
Mordell-Weil group law 
in the unHiggsed geometry. We note that,  at this point, 
we can only determine the extended Mordell-Weil group law up to vertical divisors, which will be the task of \autoref{multi_group}. This definition, however, will allow us to uniquely define a 
 \textit{generalized Shioda map} in \autoref{sec:gen_Shioda}. The latter defines divisors associated to 
massless Abelian  gauge symmetries from the generators of the postulated extended Mordell-Weil group, \textit{i.e.}~from the multi-sections. Finally, in section \autoref{sec:ExtendedMWLGT}
 we show that translations in the extended Mordell-Weil group correspond to Abelian large gauge transformations.

\subsection{A group action for fibrations with multi-sections} \label{multi_group}
  
We now present an extension of the results from the last subsection to
F-theory compactified on genus-one fibrations without section. These geometries come with multi-sections, which mark points in the elliptic fiber that are exchanged
over branch cuts in the base $B$. If they mark a set of $n$ points in the fiber, we call the multi-section an $n$-section. 
For a genus-one fibration one can always birationally
move to the Jacobian fibration, which replaces each independent $n$-section 
by a rational section and therefore constitutes an elliptic fibration.
Importantly the genus-one fibration and its Jacobian describe the same F-theory effective action 
in four or six dimensions \cite{deBoer:2001px,Braun:2014oya}. It is therefore clear that the presence of at least two 
homologically independent multi-sections indicates the existence of massless 
$U(1)$ gauge fields in the four- or six-dimensional F-theory effective field theory.\footnote{A single 
multi-section gives one massless $U(1)$ in the three- or five-dimensional effective theory, which captures 
the degree of freedom of the circle Kaluza-Klein vector.} In particular, the associated Jacobian fibration 
of a genus-one fibration with more than one multi-section will have a non-trivial Mordell-Weil group.
One can therefore ask how to identify the divisor classes associated to massless $U(1)$ gauge 
symmetries already in the genus-one fibrations. This is relevant \textit{e.g.}~for the computation of 
$U(1)$-charges or the computation of anomaly coefficients. 
Furthermore, we will argue for the existence of a group law for multi-sections.

To address these issues we first have to introduce some additional facts about 
fibrations with multi-sections and state our assumptions. 
First, recall that an $n$-section $s^{(n)}$ with divisor class 
$S^{(n)} = Div(s^{(n)})$ fulfills  
\beq
	S^{(n)}\cdot f=n\,,
\eeq
where $f$ is the class of the genus-one fiber. 
This implies that
\beq
	S^{(n)}\cdot D_\alpha\cdot D_{\beta}=n\, D^b_\alpha\cdot D^b_{\beta}\,,\qquad S^{(n)}\cdot D_\alpha\cdot D_{\beta}\cdot D_{\gamma}=n\, D^b_\alpha\cdot D^b_{\beta}\cdot D_{\gamma}^b\,,
\eeq
where the first equation applies for threefolds and the second for fourfolds. 
Second, note that it is always possible to 
find a basis of multi-sections in homology that are all of the same degree \cite{Braun:2014oya}, 
\textit{i.e.}~they cut out the same number of points in the fiber. We denote the number 
of such multi-sections by $n_{\rm ms}$ and assume $n_{\rm ms}  \geq 2$. 
We denote such a basis of $n$-sections by $s_0^{(n)},s_m^{(n)}$, $m=1,\ldots ,n_{\rm ms}-1$ 
and demand that it is minimal in the sense that there does not exist any multi-section in the geometry
that cuts out $n-1$ or fewer points.\footnote{From
now on we will always require that the considered basis of multi-sections is of this type.}
We have singled out an arbitrary multi-section which we labelled by $0$.
The divisors associated to these sections are denoted by $S_0^{(n)} = Div(s_0^{(n)})$
and $S_m^{(n)}= Div(s_m^{(n)})$ in accord with our previous notation.

To propose a group law we will work with the following assumption for genus-one fibrations throughout this section:
\begin{itemize}
\item
We assume that there exists a specialization of the complex structure of 
the fibration such that each $n$-section $s_0^{(n)}$ and $s_{m}^{(n)}$ 
splits into $n$ sections $s_0^1,\ldots, s_0^n$ and $s_m^1, \ldots , s_m^n$. 
After 
resolving the singularities in the new geometry we will denote the resulting 
space by $\hat Y_{\rm uH}$, where we indicate that this geometry captures 
the unHiggsing from a field-theoretic point of view.
In the following we impose that the rational sections $s_0^1,\ldots, s_0^n$ and
$s_m^1, \ldots , s_m^n$ are the generators of the Mordell-Weil group
supplemented by the zero section of the elliptic fibration $\hat Y_{\rm uH}$. 
We expect, however, that the following discussion can be extended to 
the more general situation in which these rational 
sections only generate a sub-lattice of the Mordell-Weil lattice.
With this simplification, the divisor homology groups of $\hat Y$ and $\hat Y_{\rm uH}$ are generated 
as follows:
\beq \label{spanH_ms}
   H_{p} (\hat Y) = \langle S_0^{(n)},  S_m^{(n)}, D_\alpha \rangle\ ,  \qquad 
   H_{p} (\hat Y) = \langle S_0^{1},\ldots, S_0^n ,S_m^1,\ldots  , S_m^{n}, D_\alpha' \rangle \ ,
\eeq
where $p=4$ for Calabi-Yau threefolds and $p=6$ for Calabi-Yau fourfolds.
Note that we will in the following assume that the theory has no non-Abelian 
gauge groups. In other words, we do not include exceptional divisors in \eqref{spanH_ms}.

\item We also introduce an \textit{unHiggsing map} $\varphi$ from the 
divisors of $\hat Y$, \textit{i.e.}~the fibrations admitting multi-sections, to the divisors 
of $\hat Y_{\rm uH}$,
\beq
	\varphi:\quad H_p (\hat{Y})\,\,\hookrightarrow \,\, H_p (\hat{Y}_{\text{uH}})\,.
\eeq
Here we have indicated that the map is injective. In addition, we require it to be an injective ring homomorphism
from the full intersection ring on $\hat{Y}$ into that of $\hat{Y}_{\text{uH}}$.
This map is defined to identify the $n$-sections with 
$n$ sections on the divisor level:
\beq \label{SplitS}
 \varphi(S_0^{(n)})=  S_0^1 + \dots + S_0^n \, ,\qquad \varphi(S_m^{(n)})= S_m^1 + \dots + S_m^n \, .
\eeq
We do not consider torsional sections in the following 
discussion. We furthermore assume that the map $\varphi$ 
acts trivially on the remaining divisors $ D_\alpha$ and is linear on the 
vector space of divisors, i.e.
\beq \label{prop_map}
    \varphi(\nu^i S_{i}^{(n)} + \nu^\alpha D_\alpha ) = 
    \nu^i S_i^1 + \dots + \nu^i S_{i}^n  + \nu^\alpha D'_\alpha   \, ,
\eeq
for some constants $(\nu^i,\nu^\alpha)$. Note that $D_\alpha$ and $D_\alpha'$ 
actually define the same divisor classes, since they both ascent form the same 
divisors in the base $B$ common to both $\hat Y$ and $\hat Y_{\rm uH}$. 
 
\end{itemize}

Note that that only a single example of a geometry with more than one independent multi-section has been studied in the
literature \cite{Klevers:2014bqa}, which is given by
an embedding of the fiber as a hypersurface into $\mathbb P^1 \times \mathbb P^1$. In these setups one finds two independent two-sections, which  do indeed
split into four sections in the prescribed way by blowing up the fiber ambient space to $dP_3$.\footnote{It is important to notice that the two toric
two-sections
of $\mathbb P^1 \times \mathbb P^1$ do not exclusively 
split into the four toric sections of $dP_3$. 
One rather has to pick four appropriate elements of the Mordell-Weil lattice 
of the blow-up that are not necessarily torically realized.}

Let us make the following preliminary ansatz for a group structure placed on the set of multi-sections
written down in homology similar to \eqref{e:MW_law}:
Choose one $n$-section $s_0^{(n)}$ as what we call the \textit{zero-$n$-section} or \textit{zero-multi-section}. 
Then two arbitrary
$n$-sections $s_1^{(n)},s_2^{(n)}$ are added according to
\begin{align}\label{e:gen_MW_law_ansatz}
 Div(s_1^{(n)} \oplus k s_2^{(n)} ) := S_1^{(n)} + k(S_2^{(n)} - S_0^{(n)} )
 + \lambda^\alpha D_\alpha  \, .
\end{align}
Making the definition \eqref{e:gen_MW_law_ansatz} precise would require to determine the 
constants $\lambda^\alpha$. However, we will argue in the following that these are not uniquely 
determined, which can be traced back to the fact that there exist divisor classes corresponding 
to genuine multi-sections that differ only in their vertical parts induced by the base homology. This implies that 
we need to talk about equivalence classes $[\cdot ]$ of divisor classes of multi-sections defined modulo vertical part. 
Furthermore, we will in the following provide evidence  that $Div(s_1^{(n)} \oplus k s_2^{(n)})$
defines a divisors class representing an actual $n$-section in the geometry when neglecting the 
vertical part. Let us stress again that our approach
just allows us to investigate how the group law for multi-section is defined in terms of homology classes.

\definecolor{blue_standard}{rgb}{0,0,255}
\begin{figure}
\begin{center}
\includegraphics[scale=0.6]{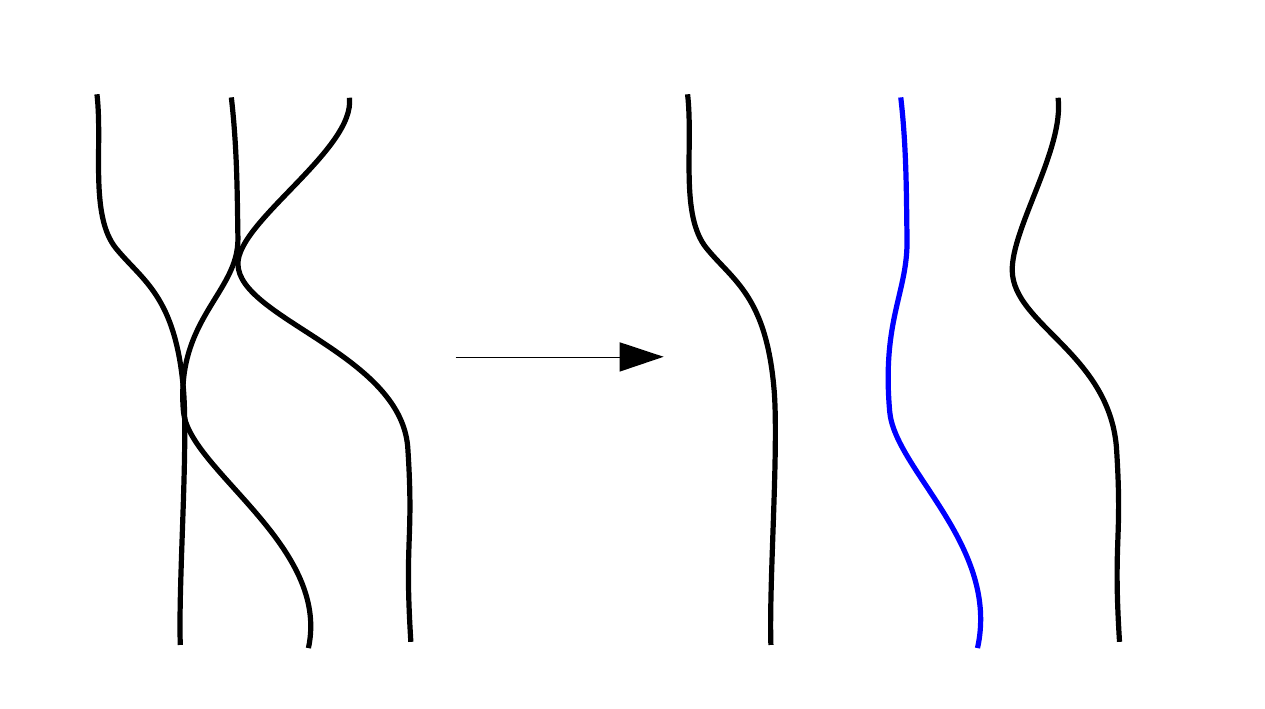}
\begin{picture}(0,0)
\put(-215,15){$s^{(3)}_0$}
\put(-105,15){$s^{1}_0$}
\put(-70,15){\textcolor{blue_standard}{$s^{2}_0$}}
\put(-45,15){$s^{3}_0$}
\put(-220,-10){zero-3-section}
\put(-90,-10){\textcolor{blue_standard}{zero-section}}
\end{picture}
\vspace*{.5cm}

\begin{minipage}{12cm}
\caption{The zero-$n$-section is chosen to contain the zero-section after unHiggsing to a setting 
with rational sections only. \label{pick-zero}}
\end{minipage}
\end{center}
\end{figure}
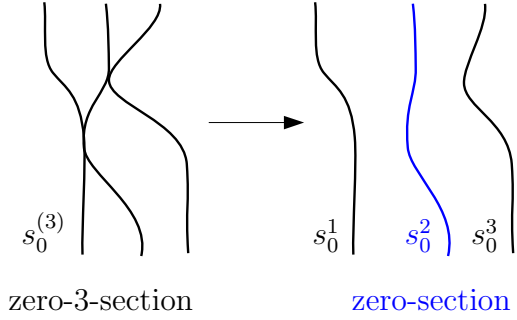

We first want to provide evidence that there is indeed 
a multi-section associated to $\tilde S^{(n)} \equiv Div(s_1^{(n)} \oplus k s_2^{(n)})$
as defined in \eqref{e:gen_MW_law_ansatz}. In order 
to do that we will check in which ways $\varphi( \tilde S^{(n)})$ 
can split into a sum of $n$ sections in the homology of $\hat Y_{\rm uH}$.
Let us denote such a set of $n$ linear independent 
sections of $\hat Y_{\rm uH}$ by $\{ \hat s^i \} $, and demand that
\begin{align}\label{e:splitting}
\varphi\big( S_1^{(n)} + k(S_2^{(n)} - S_0^{(n)} ) + \lambda^\alpha D_\alpha \big) \overset{!}{=} \sum_{i=1}^{n} \hat S^i \, ,
\end{align}
where $\lbrace \hat S^i = Div(\hat s^i)  \rbrace$ is the associated set of linearly 
independent divisors in $\hat Y_{\rm uH}$. It turns out that there are infinitely many possibilities to define an
appropriate set of sections $\lbrace \hat s^i \rbrace$.
For example, choosing one arbitrary element $s_0^l$ (for fixed $l$) as the zero-section (see \autoref{pick-zero} where \textit{e.g.}~$l=2$),
there is the very simple choice
\begin{align} \label{choice_si}
 \hat s^i := s_1^i \oplus k s_2^i \ominus k s_0^i \, ,
\end{align}
which gives the right structure \eqref{e:splitting} upon using \eqref{prop_map} and the conventional Mordell-Weil group
law \eqref{e:MW_law}. Clearly, the ansatz \eqref{choice_si}
allows us to fix the $\lambda^\alpha$ specifying the vertical part in \eqref{e:gen_MW_law_ansatz}. 
The existence of an appropriate set of $\hat s^i$ indicates that there is indeed a multi-section 
in the divisor class $Div(s_1^{(n)} \oplus k s_2^{(n)} )$, when fixing the $\lambda^\alpha$
via \eqref{e:splitting}, \eqref{choice_si} and \eqref{e:MW_law}.

\definecolor{blue_standard}{rgb}{0,0,255}
\begin{figure}
\begin{center}
\includegraphics[scale=0.4]{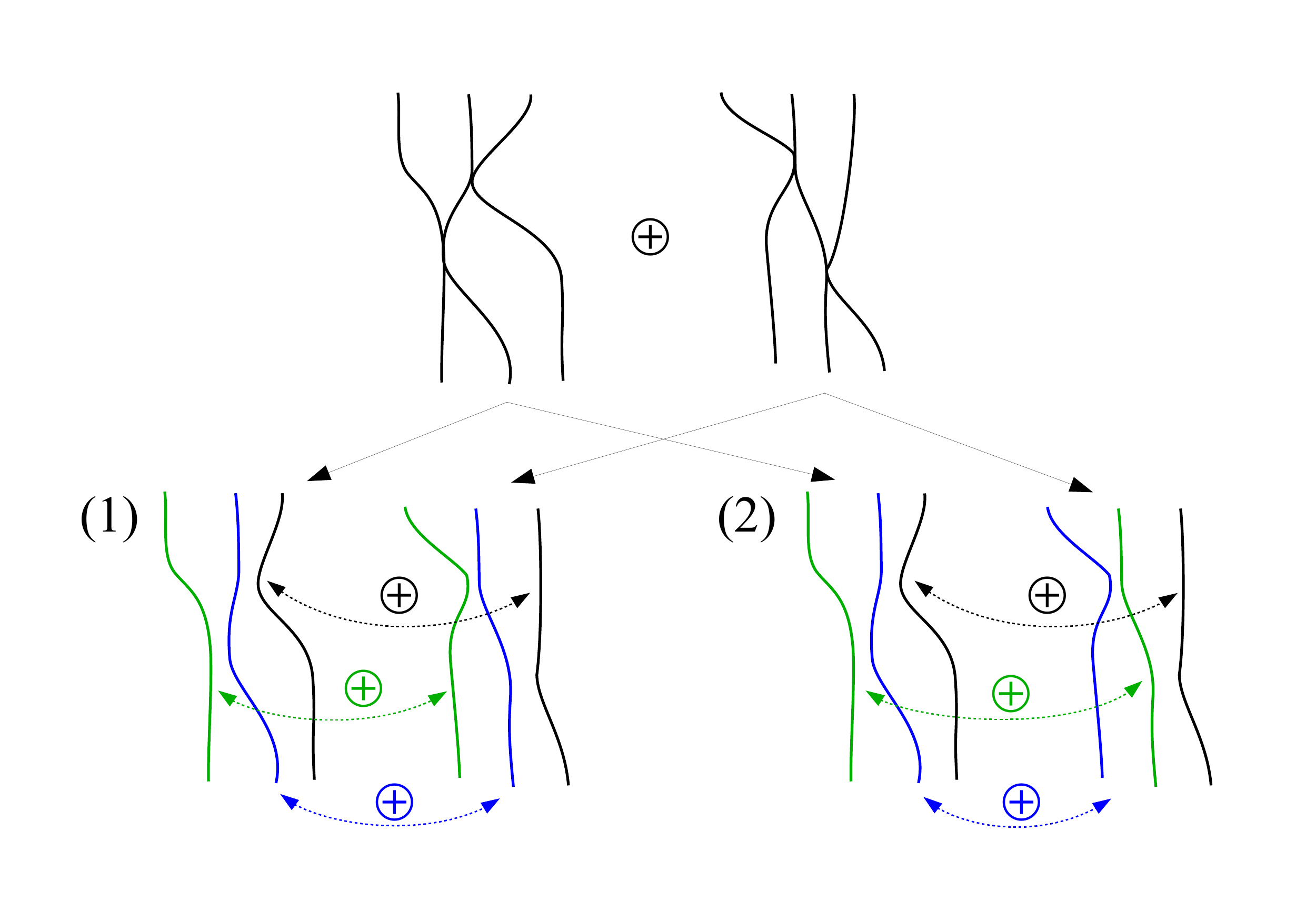}
\begin{picture}(0,0)
\put(-85,150){$\hat Y$}
\put(-10,50){$\hat Y_{\rm uH}$}
\end{picture}
\vspace*{-.5cm}

\begin{minipage}{12cm}
\caption{Moving from $\hat Y$ to the unHiggsed phase $\hat Y_{\rm uH}$ there are in general 
many ways to add the individual sections. We schematically indicate 
two choices in (1) and (2), which in general differ by their vertical parts when 
considering the associated multi-section. \label{add-sections}}
\end{minipage}
\end{center}
\end{figure}
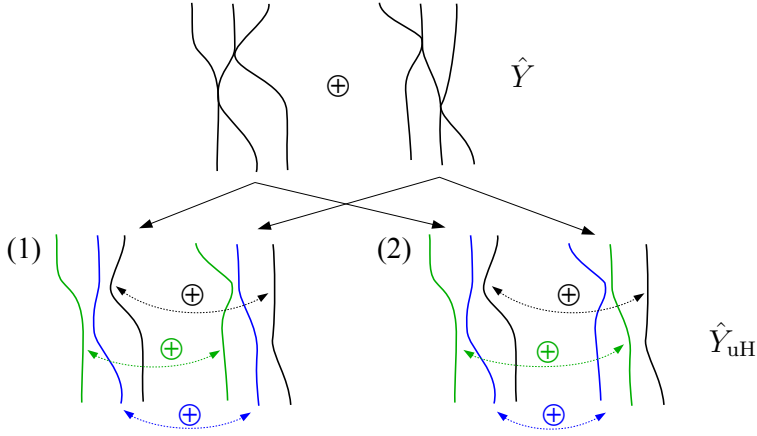

However, using merely the 
existence of sections $\hat s^i $ in $\hat Y_{\rm uH}$ satisfying \eqref{e:splitting} does not seem to fix
the class $Div(s_1^{(n)} \oplus k s_2^{(n)} )$  uniquely. In fact, 
other appropriate sets $\lbrace\hat s'^i \rbrace$ can be obtained if one picks two arbitrary sections out of
$\lbrace \hat s^i \rbrace$ and adds a third arbitrary chosen section to one of the latter while subtracting it from the other one by using the Mordell-Weil group law on $\hat Y_{\rm uH}$. Such a freedom of choice
is schematically depicted in \autoref{add-sections}. The set $\lbrace\hat s'^i \rbrace$ 
can be used to satisfy \eqref{e:splitting}, but will generally yield a different set of constants $\lambda^\alpha$
compared to the choice \eqref{choice_si}. In other words, 
the contribution from vertical divisors in \eqref{e:gen_MW_law_ansatz} is a priori not
uniquely fixed by the compatibility conditions that we impose. 
This should be contrasted with the situation for genuine sections
(\textit{i.e.}~$n=1$), in which this part is fixed by demanding
that the section squares to the canonical class in the base of the elliptic fibration. 
For multi-sections this relation to the canonical class of the base is in general not valid.\footnote{This 
can also be understood after applying the map $\varphi$ to $\hat Y$, since two of the individual 
sections $s_i^{1}$, $s_i^{2}$ arising from the $i$th $n$-section can have non-trivial intersection.}
In fact, for concrete examples
one can verify that there exist multi-sections that differ only by their vertical parts.\footnote{One can consider for example fibrations where the fiber is embedded
as a hypersurface into $\mathbb P^1 \times \mathbb P^1$. In these setups one can find four toric multi-sections from which two are independent, and the other two do indeed
differ by vertical parts from the latter in certain examples.}
Therefore, there is no immediate way that we are aware of to fix the vertical part of a multi-section. 
As stressed above this also reflects our inability to give a unique choice for the splitting \eqref{e:splitting}
and the fixation of the constants $\lambda^\alpha$ in \eqref{e:gen_MW_law_ansatz}.

It is therefore natural to define the group law \eqref{e:gen_MW_law_ansatz} only in terms of
equivalence classes of multi-sections modulo vertical divisors as
\begin{align}\label{e:gen_MW_law}
 \widehat{Div}(  [ s_1^{(n)}  ] \oplus k [ s_2^{(n)} ]   ):=   \big[ S_1^{(n)} + k(S_2^{(n)} - S_0^{(n)} ) \big]  \, .
\end{align}
In this expression we indicate that the divisors as well as the multi-sections should only 
be considered modulo vertical parts.  $ \widehat{Div}$ maps between equivalence classes 
of multi-sections and equivalence classes of divisors and reduces on representatives 
to $Div$.
Let us stress that this does not imply that one can add arbitrary vertical divisors to the right-hand side
of this equation and find an actual multi-section in the geometry. The formulation in \eqref{e:gen_MW_law}
loses some information about divisor classes supporting multi-sections. Crucially, this information  
turns out to be irrelevant in the discussion of the generalized Shioda map and therefore does not affect the 
considered physical implications of the setup.

\subsection{The generalized Shioda map} \label{sec:gen_Shioda}

In order to investigate the physical implications of the group law studied in \autoref{multi_group} 
for the effective field theory we first need to define a generalized Shioda map
for multi-sections to be defined in this section. Recall that the considered multi-sections 
were arbitrarily split as $s_0^{(n)}$, $s_{m}^{(n)}$, $m=1,\ldots,n_{\rm ms}-1$, 
where we called $s_0^{(n)}$ the zero-$n$-section.
The generalized Shioda map associates divisors $D^{(n)}_m \equiv D(s_m^{(n)})$ to 
the $n$-sections $s_m^{(n)}$. In addition, one has to define a map from $s_0^{(n)}$
to a divisor $D_{0}^{(n)}$ generalizing \eqref{e:old_base}, also to be defined below.
The Poincar\'e-dual two-forms $[D_{0}^{(n)}]$ and $[D^{(n)}_m]$ 
can then appear in the Kaluza-Klein expansion of the M-theory three-form 
$C_3$ generalizing \eqref{e:expansion_threeform} as 
\beq
   C_3 = A^\alpha \wedge [D_\alpha] + \hat A^0 \wedge [D^{(n)}_0]+  \hat A^m \wedge [D_m^{(n)}]\ ,
\eeq
where we recall that we assume the absence of exceptional divisors $D_I$ associated 
to a non-Abelian gauge group in this section. The vectors $\hat A^m$ 
correspond to $U(1)$ gauge fields in the six- or four-dimensional effective 
theory, while $\hat A^0$ will contain the degree of freedom arising from the 
 Kaluza-Klein vector along the circle when performing the F-theory 
up-lift from five/three to six/four dimensions.

If we have $n_{\rm ms} \geq 2$, as we will assume in the following, we have to define generalized Shioda maps yielding the $U(1)$ divisors $D_m^{(n)}$.
To this end, we once again borrow results from the unHiggsed geometry $\hat Y_{\rm uH}$.
As was explained in \cite{Morrison:2014era,Anderson:2014yva,Klevers:2014bqa,Garcia-Etxebarria:2014qua,Mayrhofer:2014haa,Mayrhofer:2014laa,Cvetic:2015moa}  the transition from the unHiggsed geometry $\hat Y_{\rm uH}$ 
to the genus-one fibration is described by a Higgsing in the effective
five- or three-dimensional field theory of certain matter states charged under a 
linear combination of the $n \times n_{\rm ms}$  Abelian gauge fields. 
We aim to find the proper linear combinations of these $U(1)$s 
that constitute the massless $U(1)$ vectors after the Higgsing.
To begin with we consider the divisor classes $D'_{m,l}$ on $\hat Y_{\rm uH}$ given
by 
\begin{align}\label{e:multi_div_sum}
  D'_{m,l}  : = \sum_{i=1}^{n} D_m^i - \sum_{\substack{i=1\\i\neq l}}^n D_0^i \, ,
\end{align}
where $D_m^i \equiv D(s_m^i)$, $D_0^i = D(s_0^i)$ denote the Shioda maps of $s_m^i , s_0^i$, with $i\neq l$, and we have chosen an 
element $s_0^l$ (fixed $l$) as the zero-section on $\hat Y_{\rm uH}$.
Note that the ansatz \eqref{e:multi_div_sum} is invariant under 
the exchange of all components $s_m^i , s_0^i$ of a given multi-section $s_m^{(n)}$ and the zero-multi-section $s_0^{(n)}$, respectively,
and it is consistent with the map $\varphi$ as defined in \eqref{SplitS}.
In contrast, the definition \eqref{e:multi_div_sum} of $D_{m,l}^{\prime}$ still 
depends on the choice of the zero-section $s_0^l$ and is thus not 
invariant under the exchange of components of the zero-multi-section $s_0^{(n)}$.
We therefore included an additional index $l$ in the notation. 
Inserting the explicit expressions for the Shioda maps $D_m^i , D_0^i$ and 
using \eqref{SplitS} we obtain
\begin{align}\label{e:multi_blow_up_shioda}
  D_{m,l}^{\prime}  = \varphi( S_m^{(n)} - S_0^{(n)}) - \pi \big( \varphi  (S_m^{(n)} -
 S_0^{(n)} ) \cdot S_0^l \big) \, .
\end{align}

Thus it is clear that the expression \eqref{e:multi_blow_up_shioda} is still problematic 
if one wants to move solely to the phase of the genus-one fibration, since \eqref{e:multi_blow_up_shioda}
manifestly depends on the choice
of the zero-section $s_0^l$ on $\hat Y_{\rm uH}$. However, as it follows from the discussion of \autoref{sec:ec_group_structures} and was investigated already in \cite{Grimm:2015zea}, different choices of the zero-section
are just related by large gauge transformations in the effective field theory. Therefore it seems logical to treat all phases
with different zero-section $s_0^l$ on equal footing. We thus average over all these choices and 
use again \eqref{SplitS} to obtain
\beq \label{average_D'}
   \frac{1}{n}\sum_{l} D_{m,l}^{\prime} = \varphi( S_m^{(n)} - S_0^{(n)}) - \frac{1}{n} \pi \big( \varphi  (S_m^{(n)} -
 S_0^{(n)} ) \cdot  \varphi  (S_0^{(n)})\big)\ .
\eeq
Since $\varphi$ is a ring homomorphism and all pieces lie in the image 
of $\varphi$, we can now drop the map $\varphi$ in this expression and consistently define
a generalized Shioda map $D^{(n)}_m$ for the multi-section $s_m^{(n)}$ without reference to an unHiggsed phase
\begin{align}\label{gen_Shioda}
 D^{(n)}_m :=S_m^{(n)} - S_0^{(n)} - \frac{1}{n} \pi \Big ( \big (S_m^{(n)} -
 S_0^{(n)} \big ) \cdot S_0^{(n)} \Big ) \, .
\end{align}
By construction it is evident that $U(1)$ charges $q_m$ of matter in the genus-one fibration are calculated by intersecting the associated curves
with $D^{(n)}_m$. Note that this intersection is independent of the vertical contribution in $D_m$.
Furthermore \eqref{gen_Shioda} is a generalization of the map given in \cite{Klevers:2014bqa} (without the factor $\frac{1}{n}$), where the
authors consider fibers embedded into $\mathbb P^1 \times \mathbb P^1$.
We expect that our definition of $D^{(n)}_m$ yields the correct $U(1)$ divisors in 
order to study the effective field theory of F-theory on genus-one fibrations
directly without explicit reference to an unHiggsed geometry $\hat Y_{\rm uH}$ or the Jacobian of $\hat Y$.
Exploring this effective theory in detail is however beyond the scope 
of the present paper.

Further indication that $D^{(n)}_m$ is an important object of the genus-one fibration provides the fact that
the definition \eqref{gen_Shioda} only depends on the equivalence classes $\big[S_m^{(n)}\big],\big[S_0^{(n)}\big]$.
Indeed, it is easy to check that \eqref{gen_Shioda} even provides a homomorphism from the generalized Mordell-Weil group \eqref{e:gen_MW_law}
to the Ner\'on-Severi group. Note that both of these conditions are extremely restrictive.

In a similar fashion we can construct the divisor $D_0^{(n)}$ appearing in \eqref{e:expansion_threeform}. It is the cycle that is dual to the massless linear combination
of the Kaluza-Klein
vector and a set of $n-1$ $U(1)$ vectors that are massive in the higher-dimensional theory \cite{Anderson:2014yva}. 
 These correspond to the individual constituents of the zero-multi-section under the splitting \eqref{SplitS}.
In analogy to \eqref{e:multi_div_sum} we first make the ansatz
\begin{align}\label{invariantKKsum}
 D'_{0,l} := n\,   D_0 +  \sum_{\substack{i=1\\i\neq l}}^n D_0^i \, ,
\end{align}
where $D_0^i = D (s_0^i)$ are the Shioda maps with a chosen zero-section $s_0^l$.
This expression is again invariant under the exchange of the individual sections $s_0^i$
modulo vertical divisors.
We stress that $D_0$ denotes the divisor yielding the Kaluza-Klein vector in $\hat Y_{\rm uH}$ and is therefore given as in  \eqref{e:old_base} by 
\begin{align}\label{definition_base_div}
  D_0 = S^l_0 - \frac{1}{2} \pi(S^l_0 \cdot S^l_0) \, .
\end{align}
Using in \eqref{invariantKKsum} the explicit expressions for 
the Shioda maps as well as \eqref{definition_base_div} and \eqref{SplitS} we obtain
\begin{align}
 D'_{0,l} = \varphi (S_0^{(n)}) + \frac{n}{2}K - \pi \big (\varphi (S_0^{(n)}) \cdot S_0^l \big ) \, .
\end{align}
Averaging over all zero-section choices as in \eqref{average_D'} we get
\begin{align}
 \frac{1}{n} \sum_l D'_{0,l} = \varphi (S_0^{(n)}) + \frac{n}{2}K - \frac{1}{n} \pi \big (\varphi (S_0^{(n)}) \cdot \varphi (S_0^{(n)}) \big ) \, .
\end{align}
Now we can drop the map $\varphi$ using the similar arguments as above, 
we are able to define $D_0^{(n)}$ as
\begin{align}
 D_0^{(n)} := S_0^{(n)} + \frac{n}{2}K - \frac{1}{n} \pi \big (S_0^{(n)} \cdot S_0^{(n)} \big ) \, ,
\end{align}
which is a generalization of \eqref{e:old_base}.

\subsection{Extended Mordell-Weil group and large gauge transformations} \label{sec:ExtendedMWLGT}

In this final subsection we show that, similar to the genuine Mordell-Weil group of rational sections, 
translations in the extended Mordell-Weil group lattice are in one-to-one correspondence with Abelian 
large gauge 
transformations in the effective field theory. As before,  the formulation of this group law on the 
divisor level will be completely sufficient for the question we aim to address due to the uniqueness of the generalized Shioda maps.

We begin by shifting the basis of multi-sections by $k^m$-times the $n$-section $s^{(n)}_{m}$ as
 \begin{align}
  \big[ \tilde s^{(n)}_0 \big] &:=  \big[s_0^{(n)} \big] \oplus k^m \big[ s_{m}^{(n)}\big]\, , \\
  \big[ \tilde s^{(n)}_m \big] &:= \big[s_m^{(n)}\big] \oplus k^m \big[ s_{m}^{(n)}\big]\, \, . \nn
 \end{align}
Using the group law \eqref{e:gen_MW_law} and inserting the resulting divisor classes into 
the generalized Shioda map \eqref{gen_Shioda}
then find that the transformation of $D^{(n)}_m$ is given by
\begin{align}\label{e:multi_sec_shift}
 D^{(n)}_m \mapsto D^{(n)}_m - \frac{k^p}{n} \pi \big ( D^{(n)}_{p} \cdot D^{(n)}_m\big ) \, ,
\end{align}
which differs by a factor of $\frac{1}{n}$ in the vertical part from \eqref{free_shift_Ab}.
We emphasis that in this evaluation the ambiguity in the vertical parts is 
absent after applying the generalized Shioda map.

Let us now analyze the large gauge transformations from a field theory perspective. 
Recall that in general the actual Kaluza-Klein vector mixes in the Higgsed phase with other $U(1)$s 
as dictated by the zero-multi-section $s_0^{(n)}$ \cite{Anderson:2014yva,Garcia-Etxebarria:2014qua,Mayrhofer:2014haa,Mayrhofer:2014laa,Cvetic:2015moa}. While there 
are $n-1$ massive $U(1)$s parameterized by $s_0^{(n)}$ only a single $U(1)$ remains massless.
To simplify the treatment of the large gauge transformations in such a situation, 
we again can consider the unHiggsed phase. This will allow us to show
that \eqref{e:multi_sec_shift} is induced by large gauge transformations.
In particular we consider the different splits corresponding to the divisor $D^{(n)}_{m}$. Note that $D^{(n)}_{m}$
was obtained in \eqref{average_D'} by averaging over all divisors $D_{m,l}^{\prime}$, defined in \eqref{e:multi_div_sum},
which together represent the different choices for the zero-section.
Focusing now on a particular divisor $D_{m,l}^{\prime}$ with zero-section $s_0^l$, we find that the dual gauge field in the unHiggsed phase reads
\begin{align}
 A^{\prime\,m,l} = \frac{1}{n} \Big ( \sum_{i=1}^{n} A_i^{m} - \sum_{\substack{i=1\\i\neq l}}^n A_i^0 \Big ) \, ,
\end{align}
with $A_i^0, A_i^{m}$ dual to $D_0^i, D_{m}^i$.
Our main interested is in the form of the large gauge transformation for this vector field $A^{\prime\,m}$.
Therefore let us apply large gauge transformations with winding $k^m_i$
of the individual constituents $A_{i}^{m,l}$.
We find 
\begin{align}
  A_{i}^{m} \mapsto  A_{i}^{m} - k_i^m A_l^0 \, , && A_{i}^{0} \mapsto  A_{i}^{0} \, ,
\end{align}
where $A_l^0$ denotes the Kaluza-Klein vector.
We conclude that the large gauge transformation acts on $A^{\prime\, m,l}$ as
\begin{align}
 A^{\prime\,m,l} \mapsto A^{\prime\,m,l} - \frac{\sum_{i=1}^n k^m_i}{n} A_l^{0} \, .
\end{align}
Using the results from \autoref{sec:ec_group_structures} we conclude that the dual divisors
transform as
\begin{align}
 D_{m,l}^{\prime} \mapsto  D_{m,l}^{\prime} - \frac{\sum_{i=1}^n k^p_i}{n} \pi \big ( D_{p,l}^{\prime} \cdot D_{m,l}^{\prime} \big ) \, .
\end{align}
Averaging now as in \eqref{average_D'} over the different choices for the zero-section
we can finally infer that the genuine $U(1)$ divisors $D_m^{(n)}$ in the Higgsed phase transform as
\begin{align}
  D^{(n)}_m \mapsto  D^{(n)}_m - \frac{\sum_{i=1}^n k^p_i}{n} \pi \big ( D^{(n)}_{p} \cdot D^{(n)}_m \big ) \, .
\end{align}
This is precisely what we get from \eqref{e:multi_sec_shift} for appropriate choices of the $k^m_i$.
We finally conclude that shifts in the generalized Mordell-Weil group correspond to Abelian large gauge transformations
in the Higgsed phase.

\section{Introducing an arithmetic structure on fibrations with exceptional divisors} \label{Arithmetics_nonAb}

In this section, we focus on elliptic fibrations $\hat{Y}$ with codimension-one singularities leading to
non-Abelian gauge groups with matter in F-theory. The resolution of 
singularities of the elliptic fibration at codimension-one in the base $B$ requires introducing a set 
of blow-up divisors. In \autoref{sec:defGroupStructureExeptional} we define a novel group action on the  set of these blow-up divisors  in $\hat{Y}$.
We are guided by two principles in defining this group structure, one geometric and one field theoretic one. 

First, we employ the geometric
fact that many geometries $\hat{Y}$ with a Higgsable non-Abelian gauge group can be connected by
a number of extremal transitions, corresponding to Higgsing in field theory, 
to a geometry $\hat{Y}_{\rm H}$ with a purely Abelian gauge group,  \textit{i.e.}~a number of rational sections. 
Under this transition, the Cartan $U(1)$s inside the non-Abelian gauge group are mapped to 
$U(1)$s associated to the free generators of the Mordell-Weil generators of the Higgsed 
geometry $\hat{Y}_{\rm H}$.  The postulated group structure on the blow-up divisors of the non-Abelian theory is 
then nothing but the translational symmetry in the Mordell-Weil group of the Higgsed theory 
that has been shown to be a geometric symmetry in \autoref{sec:ec_group_structures}. 
In \autoref{sec:defGroupStructureExeptional} we will assume that such a Higgs transition exists 
and exploit it to define the group structure on $\hat{Y}$. 
We show this correspondence  explicitly in the simplest case of  an adjoint Higgsing of $SU(2)$ 
to $U(1)$ in \autoref{sec:unHiggsing} 
and use induction on the number of $U(1)$s to generalize to higher rank groups. 
Thus, we see that the non-Abelian group structure is required by consistency under motion in the
moduli space of F-theory.

Second, we show in \autoref{sec:defGroupStructureExeptional} that in the  effective field theory the postulated group action manifests itself 
simply as non-Abelian large gauge transformations and is therefore trivially a symmetry
in an anomaly-free theory. Thus,
we claim that the non-Abelian group action should have a direct geometric interpretation  on $\hat{Y}$
and does generally exist for any non-Abelian setup, even those lacking Higgsings to Abelian 
theories.

We note that application of  the results from the previous
sections implies that the geometric symmetries postulated here imply the cancellation of all pure 
and mixed gauge anomalies in the effective action of F-theory compactifications on elliptically
fibered Calabi-Yau three- and fourfolds.

\subsection{A group action for exceptional divisors}
\label{sec:defGroupStructureExeptional}

As outlined at the beginning of this section, we define in the following a group structure  on the
set of resolution divisors of codimension-one singularities of an elliptic fibration $\hat{Y}$. We first motivate
the group structure geometrically by the connection between Abelian and non-Abelian
gauge groups via (un)Higgsing. Then we show that the postulated group law is identified with
non-Abelian large gauge transformations, which are automatically a symmetry of the effective
theory. Furthermore, we show that the postulated group law 
leaves key classical intersections on $\hat{Y}$ invariant; in particular the intersections of
the transformed exceptional divisors yield the same Cartan matrix as before and the transformed 
rational sections obey again the defining intersection properties of rational sections discussed 
in \autoref{Ftheoryfacts}.

We will start with a purely Abelian theory specified by an elliptic fibration $\hat{Y}_{\rm H}$ with a 
Mordell-Weil group generated by elements $s'_{m}$, $m=1,\ldots, n_{U(1)}$. 
Although the following arguments hold in general, we will assume that the Mordell-Weil group has 
no torsion elements. 
We consider an unHiggsing
to a geometry $\hat{Y}$, where a subset of the rational sections are 
turned into exceptional divisors $D_I$ corresponding to a non-Abelian gauge group $G$. 
As discussed systematically in \cite{Morrison:2014era,Cvetic:2015ioa}, such an unHiggsing is a 
tuning in the complex structure of $\hat{Y}_{\rm H}$ so that certain
rational sections coincide globally in the tuned geometry. 
Thus, $\hat{Y}$ will have a lower rank Mordell-Weil group with generators denoted by $s_n$, 
$n=1,\ldots,\tilde{n}_{U(1)}$ for $\tilde{n}_{U(1)}< n_{U(1)}$. 

We focus here on the 
simplest situation possible corresponding to a rank preserving unHiggsing,  \textit{i.e.}~a situation 
with $\text{rk}(G)=n_{U(1)}-\tilde{n}_{U(1)}$. Then
the  non-Abelian gauge theory associated to $\hat{Y}$ is Higgsed back to the 
original Abelian gauge theory specified by $\hat{Y}_{\rm H}$ by matter in the adjoint representation. 
Thus, the divisor groups of $\hat{Y}_{\rm H}$ and $\hat{Y}$ are of the same dimension and 
generated by the following elements, respectively:
\beq
	H_p (\hat{Y}_{\rm H})=\langle S'_0,S'_n,S'_I,D'_\alpha\rangle\,,\qquad
	H_p(\hat{Y})=\langle S_0,S_n,D_I,D_\alpha\rangle\,,
\eeq
where $p=4$ for Calabi-Yau threefolds and $p=6$ for Calabi-Yau fourfolds.
Here $S_0'$, $S'_n$ and $S'_I$ are divisor classes associated to the rational sections on 
$\hat{Y}_{\rm H}$, $S_0$ and $S_n$ are divisor classes  of the sections on $\hat{Y}$. 
$D_\alpha'$ and $D_\alpha$ are divisors that ascent from divisors in $B$ and 
define in fact the same classes in $\hat Y_{\rm H}$ and $\hat Y$. The 
index $I=1,\ldots,n_{U(1)}-\tilde{n}_{U(1)} $ is the same for both geometries and labels the 
sections on $\hat{Y}_{\rm H}$ that are mapped to exceptional divisors $D_I$ associated to the group
$G$ on $\hat{Y}$.

We propose that  the unHiggsing $\hat{Y}_{\rm H}\rightarrow \hat{Y}$ induces the existence of a map
$\varphi$ from the divisor group of $\hat{Y}_{\rm H}$  to that of  $\hat{Y}$,
\beq
	\varphi:\quad H_p(\hat{Y}_{\rm H})\,\,\rightarrow\,\,H_p(\hat{Y})\,,
\eeq
with certain properties to be defined next. We we will argue explicitly in \autoref{sec:unHiggsing} that the (un)Higgsing processes 
described in \cite{Morrison:2012ei,Morrison:2014era,Cvetic:2015ioa} implies the existence of a 
map $\varphi$ as described now.

We require $\varphi$ to be a bijective ring homomorphism from the full intersection ring
on $\hat{Y}_{\rm H}$ to that on $\hat{Y}$, \textit{i.e.}~to commute with the intersection pairing of 
divisors and to be linear. The image of $\varphi$ on the generators of $H_p(\hat{Y}_{\rm H})$  with $p=4$ (or $p=6$ for fourfolds)  is given by
\beq \label{eq:unHiggsing}
\varphi( S_0')= S_0 \,,\quad
\varphi(S_{n}')= S_{n} \, , \quad \varphi(D'_{I})=D_{I}\,, \quad \varphi(D'_{\alpha})=D_{\alpha}\,.
\eeq
We emphasize that $\varphi$ maps the  Shioda map $D'_I$ of the rational section $s_I'$ 
to a Cartan divisor $D_I$ of the unHiggsed gauge group $G$ on 
$\hat{Y}$. Note, however, that 
\eqref{eq:unHiggsing} implies that $\varphi$ does \textit{not}  map the Shioda map $D'_n$ of 
a section $s'_n$ on $\hat{Y}_{\rm H}$ to the 
Shioda map $D_n$ of $s_n$ on $\hat{Y}$. This is clear as the formula for  $D_n$, 
according to \eqref{e:old_shioda}, involves the Cartan divisors on $\hat{Y}$ that are 
absent on $\hat{Y}_{\rm H}$ and, consequently, do not appear in the formula for $D_n'$.

We are now in the position to investigate the image of a translation in the Mordell-Weil group
of $\hat{Y}_{\rm H}$ under the map $\varphi$ to the unHiggsed geometry $\hat{Y}$.
We are particularly interested in shifts by rational sections $s'_I$, whose associated 
Shioda maps $D_I'$ map to Cartan divisors $D_I$ in $\hat{Y}$.
To this end we recall the action of a Mordell-Weil translation on $\hat{Y}_{\rm H}$ on its divisor group.
First, we express the Mordell-Weil 
translations on $\hat{Y}_{\rm H}$  conveniently in terms of the $D'_I$. 
Shifting the Mordell-Weil lattice on $\hat{Y}_{\rm H}$
by a vector $\oplus k^{ I} s'_{I}$ we rewrite
\eqref{e:MW_law} for all sections $s'_{\cM} :=\{ s'_0, s'_m\}$ as
\begin{align} \label{eq:GroupLawDm}
 Div(s'_{\cM}\oplus k^{ I} s'_{I})  = S'_{\cM} + \sum_{I}k^{I}D'_{I} -
 \frac{1}{2}\sum_{I, J}k^{I} k^{J} \pi(D'_{I} \cdot D'_{J})
 - \sum_{ I}k^{ I} \pi(S'_\cM \cdot D'_{I}) \, .
\end{align}
We also recall the general Mordell-Weil group action on a Shioda map $D'_m$ of a section 
$s'_m$ as given in 
\eqref{free_shift_Ab}.
We now perform the unHiggsing by applying the ring homomorphism $\varphi$, employing
\eqref{eq:unHiggsing}, to the formulae in
\eqref{eq:GroupLawDm} and \eqref{free_shift_Ab}. We find the 
following transformation of divisor classes of sections and Cartan divisors on 
$\hat{Y}$ by lifting the Mordell-Weil translations on $\hat{Y}_{\rm H}$:
\begin{subequations} \label{eq:DsZeroNodeShift}
\begin{align} 
 \tilde S_{0} &= S_{0} + \sum_{I}k^{I}D_I - \frac{1}{2}\sum_{I,J}k^I k^J \pi(D_I \cdot D_J) \\
 \tilde S_{n} &= S_{n} + \sum_{I}k^{I}D_I - \frac{1}{2}\sum_{I,J}k^I k^J \pi(D_I \cdot D_J)
 - \sum_{I}k^{I} \pi( {S}_n \cdot D_I) \\
  \tilde{D}_I &= D_I - \sum_{J}k^{J} \pi (D_J \cdot D_I) \, .
\end{align}
\end{subequations}
We note that the map $\varphi$ simply amounts to 
$\sum_{I} k^{I}D'_{I} \mapsto \sum_{I}k^{I}D_I$, as follows from \eqref{eq:unHiggsing}. 
Then, we have additionally used $\pi(S_0\cdot D_I)=0$  in the first equation since the zero 
section does not pass through the Cartan divisors on $\hat{Y}$ as expected and $\pi(D_n\cdot D_I)=0$ by definition of the Shioda map on the unHiggsed geometry 
$\hat{Y}$.

From a field theory point of view it is clear that the shifted classes \eqref{eq:DsZeroNodeShift} 
correspond to non-Abelian large gauge transformations along the Cartan subalgebra.
Indeed one finds that under \eqref{eq:DsZeroNodeShift} the divisors on $\hat{Y}$
transform as 
\begin{align} \label{D-transformZeroNode}
 \begin{pmatrix}
  \tilde D_{0}\\[8pt]
  \tilde D_{I}\\[8pt]  
  \tilde D_{n}\\[8pt]
  \tilde D_{\alpha}
 \end{pmatrix} = 
\begin{pmatrix*}[c]
 1 &  k^J & 0  & \frac{k^J k^L}{2} \cC_{JL} b^\beta \\[8pt]
0 & \delta^{J}_{I} & 0 &   k^K \cC_{IK} b^\beta \\[8pt]
  0 & 0 & \delta_n^k & 0 \\[8pt]
 0 & 
 0
  & 0& \delta^\beta_\alpha
\end{pmatrix*}\cdot
\begin{pmatrix}
  D_{0} \\[8pt]
  D_{J} \\[8pt]
  D_{k} \\[8pt]
  D_{\beta} \, 
 \end{pmatrix}\,,
\end{align}
where we have used \eqref{evaluate_DD} and again recall that $\pi(D_n\cdot D_I)=0$.
Indeed, \eqref{D-transformZeroNode} is precisely  the formulae for a non-Abelian large 
gauge transformation given in \eqref{eq:LGTgeneral_1} along the non-Abelian Cartan gauge 
field $\sum_{J}k_{J}A^J$, {i.e.}~for $k^m=0$, so that the combination 
$C_3=A^\Lambda\wedge [D_\Lambda]$ remains invariant. 

In the following we will impose these shifts in F-theory compactifications with non-Abelian gauge 
symmetry independently of an existing
adjoint Higgsing to the maximal torus of $G$. 
We conclude by showing that the transformed divisor
classes \eqref{eq:DsZeroNodeShift} on $\hat{Y}$
obey the key properties \eqref{e:shioda_orth} and \eqref{blowup-cond} of new Cartan divisors 
and new rational sections, respectively, so that the gauge algebra and the rank of the 
Mordell-Weil group are invariant. First of all let us note that the $\tilde S_{\cM} = \{\tilde S_{0}, \tilde S_{n}\}$ define good divisor classes for sections. Indeed we find that
\begin{align}
 \tilde S_{0} \cdot \tilde D_I &= 0 \\
  \tilde S_{\cM} \cdot \cE &= 1 \\
  \pi (\tilde S_{\cM} \cdot \tilde S_{\cM} ) &= K \, .
\end{align}
Second we check that also the other classical intersection numbers such as \eqref{evaluate_DD} 
for the divisors $\tilde D_I$ are not changed.  This indicates that there might exist a new geometry interpretation
of the transformed divisors $(\tilde S_0 ,\tilde S_n,\tilde D_I)$ 
as sections and exceptional divisors in an associated geometry.
It also hints to the existence of a geometric interpretation for the Mordell-Weil translations
lifted from $\hat{Y}_{\rm H}$ to $\hat{Y}$.

To close this subsection, let us note that we can push the analogy to the elliptic fibration 
with rational sections even further by defining a so-called \textit{zero-node} $\Sigma_0$ (see also \cite{Grimm:2015zea}).
We introduce $\Sigma_0$  as
\begin{align}
 \Sigma_0 := \sum_{I}k^{I}D_I\ .
\end{align}
Using this definition the transformations \eqref{eq:DsZeroNodeShift} can be rewritten in a simpler 
form eliminating all $k^I$-dependence. The freedom to make a shift by a large 
non-Abelian gauge transformation then translates to `picking a zero-node'. 
This is analogous to picking a zero-section in examples with multiple rational sections.
For the latter case it has been argued in \cite{Grimm:2015zea} that this should constitute 
an actual symmetry of the M-theory to F-theory limit and therefore implies cancellation 
of Abelian anomalies. For the non-Abelian large gauge transformations and 
the group action introduced here such a geometric symmetry principle has yet 
to be established, but would guarantee the cancellation of all non-Abelian anomalies.

 \subsection{Arithmetic group structures from Higgs transitions}
 \label{sec:unHiggsing}
 
Given a gauge theory with non-Abelian gauge group $G$ and matter in the adjoint 
representation, we can Higgs to $U(1)^{r}$ with $r=\text{rk}(G)$ by switching on VEVs
along the Cartan generators in the adjoint. The inverse process is called unHiggsing a $U(1)$ 
symmetry. Various examples of unHiggsing $U(1)$ symmetries in F-theory have been considered,
see {e.g.}~the most recent works 
\cite{Morrison:2012ei,Morrison:2014era,Klevers:2014bqa,Cvetic:2015ioa} on 
unHiggsings of up to two $U(1)$s. We will 
employ the unHiggsing of $U(1)$ symmetries to non-Abelian groups in the following in order to 
provide further geometrical evidence for the existence of the group structure on the
exceptional divisors postulated in \autoref{sec:defGroupStructureExeptional}. For simplicity, 
we focus here on the simplest case of the unHiggsing of one $U(1)$ to $SU(2)$ as studied in
\cite{Morrison:2012ei,Morrison:2014era}. By induction over the number of $U(1)$s, as 
suggested in \cite{Cvetic:2015ioa}, the obtained results are expected to generalize to higher
rank non-Abelian gauge groups.

It has been shown by Morrison and Park in \cite{Morrison:2012ei} that the normal form of a 
general elliptic fibration with a rank one Mordell-Weil group, \textit{i.e.}~a single $U(1)$, is a 
Calabi-Yau hypersurface $\hat{Y}_H$ with elliptic fiber given as  the quartic hypersurface in 
the blow-up of $\mathbb{P}^2(1,1,2)$, denoted $\text{Bl}_1\mathbb{P}^2(1,1,2)$. 
This space has a toric description.
Denoting the projective coordinates  on $\text{Bl}_1\mathbb{P}^2(1,1,2)$ by $[u:v:w:e]$, 
where $e=0$ is exceptional divisor of the blow-up (with map $[u:v:w:e]\mapsto [ue:v:we]$ to 
$\mathbb{P}^2(1,1,2)$), the elliptic fibration can be brought into the form\footnote{Note that the coefficient of $ew^2$ is set to one in order to avoid the $\mathbb{Z}_2$-
singularity at $u=v=0$, which  would give rise to a codimension-one singularity of type $I_2$, 
{i.e.}~an SU(2) gauge group in F-theory, see \cite{Anderson:2014yva,Klevers:2014bqa}
for an analysis of the geometry with this additional singularity.}
\beq \label{eq:resolvedQuartic}
	e w^2+bv^2w=u(c_0u^3e^3+c_1u^2e^2v+c_2uev^2+c_3v^3)\,.
\eeq
The coefficients $c_i$, $i=0,1,2,3$, are sections in specific line bundles that are determined by 
the the requirement that \eqref{eq:resolvedQuartic} defines a well-defined section of a line 
bundle on the base $B$ and obeys the Calabi-Yau condition:
\beq
\text{
\begin{tabular}{c|c}
\text{Section} & \text{Class}\\
\hline
	$[c_0]$&$-4K-2[b]$\rule{0pt}{13pt} \\
	$[c_1]$&$-3K-[b]$\rule{0pt}{12pt} \\
	$[c_2]$&$-2K$\rule{0pt}{12pt} \\
	$[c_3]$&$-K+[b]$\rule{0pt}{12pt} \\
	$[b]$&$[b]$\rule{0pt}{12pt} \\
\end{tabular}
}
\eeq
Here, we denote the divisor class of a section by $[\cdot]$ and $-K$ is the anti-canonical 
divisor of $B$. Note that the class $[b]$ of the divisor  $b=0$ is a free parameter of the 
Calabi-Yau manifold $\hat{Y}$. 

The two rational sections of the elliptic fibration are given by
\beq \label{eq:ratSections}
	s'_0:\quad [0:1:1:-b]\,,\qquad s'_1:\quad  [b:1:c_3:0]\,,
\eeq
where we picked $s_0$ as the zero section.\footnote{This convention deviates from the 
one chosen in \cite{Morrison:2012ei}, but is physically equivalent \cite{Grimm:2015zea}.} The Shioda map \eqref{e:old_shioda} of the section $s'_1$ reads
\beq
\label{eq:ShiodaSigma1}
D'_1=S'_1-S'_0+K-[b]=S'_1-S'_0-[c_3]\,,
\eeq
where we denoted the  homology class of the two sections by $S'_0=Div(s'_0)$ and $S'_1=Div(s'_1)$.

The divisor $D'_1$ supports the $U(1)$ of the F-theory compactification 
on \eqref{eq:resolvedQuartic} as can be seen from the expansion \eqref{e:expansion_threeform}
of the M-theory three-form. Shifting the origin in the Mordell-Weil lattice, as discussed in 
\autoref{sec:FreeMWShifts}, yields the new divisor classes \eqref{free_shift_Ab} that were 
shown to correspond to large Abelian gauge transformations.

The unHiggsing of the $U(1)$ to an $SU(2)$ is performed by tuning 
$b\mapsto 0$ in the elliptic fibration \eqref{eq:resolvedQuartic}, as discussed in \cite{Morrison:2012ei,Morrison:2014era}, so that the rational sections in \eqref{eq:ratSections} coincide 
globally, except for the locus $c_3=0$. As the fiber is toric, this can be achieved by blowing up 
at $u=e=0$, which amounts to replacing
\beq \label{eq:buQuartic}
	u\mapsto ue_1\,,\qquad e\mapsto ee_1\,,
\eeq
where $e_1=0$ is a new exceptional divisor. 
The hypersurface equation for the unHiggsed geometry $\hat{Y}$
after blow-up reads
\beq \label{eq:resolvedQuartic1}
	e w^2=u(c_0u^3e^3e_1^6+c_1u^2e^2e_1^4v+c_2uee_1^2v^2+c_3v^3)\,.
\eeq

The single remaining section on $\hat{Y}$, now denoted by 
$s_0$, is described by $e_1=0$ after blow-up, with 
coordinates $[u:v:w:e:e_1]$ reading
\beq
	s_0:\quad [1:1:1:c_3:0]\,
\eeq
showing that $s_0$ is holomorphic.
We note that the blown-up hypersurface has a Kodaira singularity of type $I_2$ at $c_3=0$ 
corresponding to an $SU(2)$ gauge group in F-theory.  Indeed, by setting $c_3=0$ in \eqref{eq:resolvedQuartic1} 
we obtain 
\beq
	e (w^2-u(c_0u^3e^3e_1^6+c_1u^2e^2e_1^4v+c_2uee_1^2v^2)=0\,,
\eeq
which describes two $\mathbb{P}^1$'s intersecting at two points. Thus, we identify $S^{SU(2)}=\{c_3=0\}$ as the divisor supporting the $SU(2)$ gauge group.  
As the zero section $s_0$ passes through the $\mathbb{P}^1$ given by $e=0$, we 
determine the class of the Cartan divisor $D_1$ as 
\beq
	D_1=[c_3]-[e]\,.
\eeq
Furthermore, we see that the divisor $u=0$ does not intersect the hypersurface 
\eqref{eq:resolvedQuartic1}, {i.e.}~$\hat{Y}\cap \{u=0\}=0$, due to the Stanley-Reissner ideal of the blown-up ambient space.
Using these observations, we infer that the pull-back of the Shioda map \eqref{eq:ShiodaSigma1}
of the original rational section $s'_1$ to the  unHiggsed geometry $\hat{Y}$ 
reads
\beq
	D_1'\mapsto [e]-[c_3]=-D_1\,.
\eeq
Clearly, we have $S_0'\mapsto S_0$ while vertical divisor $D_\alpha$ map trivially.
These are, up to the irrelevant sign in the map of $D_1'$, precisely the properties of the map $\varphi$ defined in
\eqref{eq:unHiggsing}.

In summary, we see that the Shioda map of the rational sections is mapped, up to sign, to the 
Cartan divisor of the unHiggsed $SU(2)$ gauge group on $\hat{Y}_{\text{uH}}$. Consequently, the Mordell-Weil shift of a 
rank one Mordell-Weil group, as introduced in \autoref{sec:FreeMWShifts}, is mapped under 
the transition corresponding to the unHiggsing to $SU(2)$ to a similar shift of
divisors, where the  Shioda map is replaced by the Cartan divisor of the 
$SU(2)$ (the sign can be absorbed by the integer $k$ in \eqref{eq:MWshift}). In addition, a 
similar replacement should apply for unHiggsing a higher rank Mordell-Weil group by induction on 
its rank, as discussed in \cite{Cvetic:2015ioa}. This is expected to
establish the existence of the group law postulated in \autoref{sec:defGroupStructureExeptional}
on the Cartan divisors of any non-Abelian gauge group 
in F-theory that can be Higgsed in an adjoint Higgsing to a purely Abelian gauge group. We 
propose that this group law exists even for those non-Abelian groups that can not be Higgsed,
such as the non-Higgsable clusters in \cite{Morrison:2012np}.

\section{Conclusions}

In this work, we have systematically studied the relationship between the F-theory effective 
theory on a circle and the geometry of resolved elliptic fibrations.  We have developed for the 
first time the detailed dictionary between large gauge transformations along
the circle direction and arithmetic structures as well as  geometric symmetries of the elliptic fibration. 
Our discussion is model-independent, \textit{i.e.}~the structures and correspondences we have 
uncovered are present in any elliptic or genus-one fibration suitable for F-theory.
We summarize the key results of this work and point out some avenues for future works
in the following.

We have shown, as suggested in \cite{Grimm:2015zea}, that the translational symmetry in 
the Mordell-Weil lattice of rational sections (the so-called translation-by-$Q$ map) 
corresponds to a combination of Abelian and 
fractional non-Abelian large gauge transformations. This completes the proof of cancellation of all 
Abelian gauge anomalies  in F-theory of \cite{Grimm:2015zea}. 
We have explicitly provided the change of all 
divisor classes under a translation by an arbitrary combination of free generators of the 
Mordell-Weil group. 
Most notably, we find that one has to combine the Abelian large gauge transformation
with an additional 
fractional non-Abelian large gauge transformation if there exist fractional charges. 
Geometrically, this link follows due to fractional contributions to the Shioda map of a rational 
section, which lead to fractional charges. Field-theoretically, fractional charges are possible 
in the presence of non-Abelian groups and corresponding Abelian large gauge 
transformations are then only well-defined if augmented by appropriate 
fractional non-Abelian ones.
We emphasize that the Mordell-Weil translation also acts non-trivially on exceptional divisors.  
While the geometric explanation of this fact is beyond the scope of this work, we expect
a better understanding of the Mordell-Weil group law at singular or resolved fibers at codimension-one to shed light on this issue.

We have also extended the translational symmetry in the Mordell-Weil lattice to shifts
by its torsional elements. Similar as before, we have shown that these transformations 
correspond to special fractional non-Abelian large gauge transformations along the circle. 
Again, we have presented to action on all divisor classes in the geometry where we stress 
the non-trivial action on Cartan divisors. 

Extending the correspondence between the arithmetic structure of the Mordell-Weil group and symmetries 
in the effective theories of F-theory,
we have found compelling evidence for the existence of a group structure on the set of multi-sections in 
genus-one fibrations. We refer to this new group as the \textit{extended Mordell-Weil group}.
The study of this novel structure on the divisor level allows us to define a  \textit{generalized Shioda map} 
for multi-sections, as suggested in \cite{Klevers:2014bqa}, which can be used \textit{e.g.}~for the computation
of U(1)-charges of matter.
The guiding principle in formulating the group law is, given a geometry with a $n$-section, the 
existence of a different geometry with $n$ sections related by unHiggsing, \textit{i.e.}~a number 
of conifold transitions. The group law on the multi-sections is then merely a lift of parts of the 
group law on the sections in the unHiggsed geometry. Using the other results of this work 
mentioned before, this implies that translations in the extended Mordell-Weil group correspond to 
Abelian large gauge transformations in the unHiggsed theory. While we have introduced 
the extended Mordell-Weil group for torus fibrations without rational sections, we speculate that 
there equally exists an extended group structure for multi-sections in geometries
also admitting rational sections. 

It is an interesting problem for future works to formulate the postulated
group law of the extended Mordell-Weil group in terms of operations on the coordinates of 
multi-sections, similar to the usual group law for rational points on elliptic curves. This would 
reveal the underlying geometric symmetries of genus-one and their associated Jacobian fibrations
more directly.

Finally, we have argued for the existence of a novel group structure on the set of exceptional
divisors for resolved geometries with non-Abelian gauge groups in  F-theory. Here, we have been 
guided by the effect of non-Abelian large gauge transformations along the circle on the 
lower-dimensional effective theory on the Coulomb branch. We have discovered a group
structure and formulated its action on divisor classes of the resolved geometry. 
The resulting new divisors classes are shown to have all the properties necessary to be 
interpreted as  new exceptional divisors and rational sections.
The  transformations found are formally completely analogous to those induced by translations in 
a Mordell-Weil lattice. Indeed, we argue that this is expected as any non-Abelian gauge theory 
with adjoint matter can be Higgsed in a rank preserving way to an Abelian theory. We show that 
under this transition the  postulated group law on exceptional divisors is mapped to 
the group law arising from translations in the Mordell-Weil group of the geometry 
yielding the Abelian theory. We have demonstrated this 
explicitly for a single $U(1)$ described by the geometry in \cite{Morrison:2012ei,Morrison:2014era} and used 
induction on the number of $U(1)$s, following \cite{Cvetic:2015ioa}, to generalize to higher rank 
groups.  Moreover, as its existence relies just on the existence of 
non-Abelian large gauge transformations, we expect the postulated group structure on 
exceptional divisors to always be realized as a geometric symmetry for resolved elliptic fibration 
with codimension-one singularities.

The explicit construction of the geometric action corresponding to non-Abelian large gauge 
transformations on the resolved elliptic fibration remains an open problem worth
being addressed in future works. We speculate that the relevant 
geometric operations involve an appropriate combination of base change and quadratic twist. 
Having an explicit realization of this geometric symmetry action would provide a general proof
of cancellation of all mixed and pure gauge anomalies in any given F-theory compactification.


 \subsubsection*{Acknowledgments}
We would like to thank Andreas Braun, 
Mirjam Cvetic, Antonella Grassi, Jan Keitel, Noppadol Mekareeya, Dave Morrison, Jonas Reuter, Sakura Sch\"afer-Nameki, and Wati Taylor for illuminating discussions. 
This work was supported by a grant of the Max Planck Society. It was in part 
prepared at the Aspen Center for Physics, which is supported by National Science 
Foundation grant PHY-1066293. D.K.~thanks the Max-Planck Institute for Physics in Munich for hospitality during completion of this project.

\appendix

\section{Lie theory conventions}\label{app:Lieconventions}

In this appendix we summarize our conventions for the Lie algebra theory used in 
this work.

We consider a simple Lie algebra $\mathfrak{g}$ associated to the Lie group $G$. 
The definition of a (preliminary) basis of Cartan generators $\lbrace\tilde T_I \rbrace$ with 
\begin{align}
 \tr_{\rm f} ( \tilde T_I \tilde T_J ) = \delta_{IJ} \ ,
\end{align}
will allow us to fix the normalization of the root lattice. The trace $\tr_{\rm f}$ is taken in the fundamental representation.
As in \autoref{sec:gensetup+reduction} we denote the simple roots by $\boldsymbol{\alpha}_I$, $I=1,\dots ,\rk \mathfrak{g}$, the simple coroots
are denoted by $\boldsymbol{\alpha}_I^\vee := \frac{2 \boldsymbol{\alpha}_I}{\langle \boldsymbol{\alpha}_I , \boldsymbol{\alpha}_I \rangle}$.

In order to match with the geometric setup it is important to introduce a 
coroot-basis $\lbrace T_I \rbrace$ for the Cartan-subalgebra. It is defined by
\begin{align}
 T_I := \frac{2\,\boldsymbol{\alpha}_I^i \tilde T_i}{\langle\boldsymbol{\alpha}_I ,\boldsymbol{\alpha}_I \rangle}
\end{align}
with $\boldsymbol{\alpha}_I^i$ the components of the simple roots.
We furthermore define the (normalized)
coroot intersection matrix $\cC_{IJ}$ as
\begin{align}\label{e:def_coroot_int_mat}
 \cC_{IJ} = \lambda_G^{-1} \langle \boldsymbol{\alpha}^\vee_I , \boldsymbol{\alpha}^\vee_J \rangle \ ,
\end{align}
with
\begin{align}
 2 \lambda_G^{-1} = \langle \boldsymbol{\alpha}_{\textrm{max}}, \boldsymbol{\alpha}_{\textrm{max}} \rangle  \  ,
\end{align}
where $\boldsymbol{\alpha}_{\textrm{max}}$ is the root of maximal length.
The normalization of the Cartan generators $T_I$ (in the coroot basis) is then given by
\begin{align}
 \tr_{\rm f} (T_I T_J) = \lambda_G \, \cC_{IJ} \, .
\end{align}
Furthermore for some weight $w$ the Dynkin labels are defined as
\begin{align}
 w_I := \langle \boldsymbol{\alpha}^\vee_I , w \rangle \, .
\end{align}
The Coxeter labels $a^I$ denote the components of the highest root $\theta$ in the expansion
\begin{align}
 \theta =: \sum_{I} a^I \boldsymbol{\alpha}_I \, .
\end{align}
Finally in \autoref{t:lie_conventions} we display the numbering of the nodes in the Dynkin diagrams, the Coxeter labels and the definition of the fundamental
representations of all simple Lie algebras, as well as the values for the normalization factors $\lambda_G$ in our conventions.
\begin{table}[H]
\begin{center}
\begin{tabular}{|m{1.2cm}||m{5.3cm}|m{3.2cm}|m{3.7cm}|m{0.8cm}|}
\hline
algebra & Dynkin diagram &  Coxeter labels & fund. representation & $\lambda_G$\\
\hline \hline
$A_n$ & \scalebox{3}{\begin{dynkin}\tikzstyle{every node}=[font=\tiny\tiny]
    \node[align=left, scale=0.5] at (\dynkinstep*1,-.15cm){1};
    \node[align=left, scale=0.5] at (\dynkinstep*2,-.15cm){2};
    \node[align=left, scale=0.5] at (\dynkinstep*3,-.15cm){3};
    \node[align=left, scale=0.5] at (\dynkinstep*6,-.15cm){n-1};
    \node[align=left, scale=0.5] at (\dynkinstep*7,-.162cm){n};
    \node[align=left, scale=0.5] at (\dynkinstep*1,.1cm){};
    \dynkinline{1}{0}{4}{0};
    \dynkindots{4}{0}{5}{0};
    \dynkinline{5}{0}{7}{0};
    \foreach \x in {1,2,3,6,7}
    {
       \dynkindot{\x}{0}
    }
  \end{dynkin}} & $(1,1,1,\dots,1,1)$ & $(1,0,0,\dots,0,0)$ & 1\\
\hline 
$B_n$ & \scalebox{3}{\begin{dynkin}\tikzstyle{every node}=[font=\tiny]
    \node[align=left, scale=0.5] at (\dynkinstep*1,-.15cm){1};
    \node[align=left, scale=0.5] at (\dynkinstep*2,-.15cm){2};
    \node[align=left, scale=0.5] at (\dynkinstep*3,-.15cm){3};
    \node[align=left, scale=0.5] at (\dynkinstep*6,-.15cm){n-1};
    \node[align=left, scale=0.5] at (\dynkinstep*7,-.162cm){n};
    \node[align=left, scale=0.5] at (\dynkinstep*1,.1cm){};
    \dynkinline{1}{0}{4}{0};
    \dynkindots{4}{0}{5}{0};
    \dynkinline{5}{0}{6}{0}
    \dynkindoubleline{6}{0}{7}{0};
    \foreach \x in {1,2,3,6,7}
    {
       \dynkindot{\x}{0}
    }
  \end{dynkin}} & $(1,2,2,\dots,2,2)$ & $(1,0,0,\dots,0,0)$ & 2\\ 
\hline 
$C_n$ & \scalebox{3}{\begin{dynkin}\tikzstyle{every node}=[font=\tiny]
    \node[align=left, scale=0.5] at (\dynkinstep*1,-.15cm){1};
    \node[align=left, scale=0.5] at (\dynkinstep*2,-.15cm){2};
    \node[align=left, scale=0.5] at (\dynkinstep*3,-.15cm){3};
    \node[align=left, scale=0.5] at (\dynkinstep*6,-.15cm){n-1};
    \node[align=left, scale=0.5] at (\dynkinstep*7,-.162cm){n};
    \node[align=left, scale=0.5] at (\dynkinstep*1,.1cm){};
    \dynkinline{1}{0}{4}{0};
    \dynkindots{4}{0}{5}{0};
    \dynkinline{5}{0}{6}{0}
    \dynkindoubleline{7}{0}{6}{0};
    \foreach \x in {1,2,3,6,7}
    {
       \dynkindot{\x}{0}
    }
  \end{dynkin}} & $(2,2,2,\dots,2,1)$ & $(1,0,0,\dots,0,0)$ & 1\\
\hline 
$D_n$ & \scalebox{3}{\begin{dynkin}\tikzstyle{every node}=[font=\tiny]
    \node[align=left, scale=0.5] at (\dynkinstep*1,-.15cm){1};
    \node[align=left, scale=0.5] at (\dynkinstep*2,-.15cm){2};
    \node[align=left, scale=0.5] at (\dynkinstep*3,-.15cm){3};
    \node[align=left, scale=0.5] at (\dynkinstep*6,-.15cm){n-2};
    \node[align=left, scale=0.5] at (\dynkinstep*7,-.35cm){n-1};
    \node[align=left, scale=0.5] at (\dynkinstep*7,.35cm){n};
    \dynkinline{1}{0}{4}{0};
    \dynkindots{4}{0}{5}{0};
    \dynkinline{5}{0}{6}{0}
    \foreach \x in {1,2,3,6}
    {
       \dynkindot{\x}{0}
    }
    \dynkindot{7}{.8}
    \dynkindot{7}{-.8}
    \dynkinline{6}{0}{7}{.8}
    \dynkinline{6}{0}{7}{-.8}
  \end{dynkin}} & $(1,2,2,\dots,2,1,1)$ & $(1,0,0,\dots,0,0,0)$ & 2\\
\hline 
$E_6$ & \scalebox{3}{\begin{dynkin}\tikzstyle{every node}=[font=\tiny]
    \node[align=left, scale=0.5] at (\dynkinstep*1,-.15cm){1};
    \node[align=left, scale=0.5] at (\dynkinstep*2,-.15cm){3};
    \node[align=left, scale=0.5] at (\dynkinstep*3,-.15cm){4};
    \node[align=left, scale=0.5] at (\dynkinstep*4,-.15cm){5};
    \node[align=left, scale=0.5] at (\dynkinstep*5,-.15cm){6};
    \node[align=left, scale=0.5] at (\dynkinstep*3.5,.25cm){2};
    \foreach \x in {1,...,5}
    {
        \dynkindot{\x}{0}
    }
    \dynkindot{3}{1}
    \dynkinline{1}{0}{5}{0}
    \dynkinline{3}{0}{3}{1}
  \end{dynkin}} & $(1,2,2,3,2,1)$ & $(0,0,0,0,0,1)$ & 6\\
\hline 
$E_7$ & \scalebox{3}{\begin{dynkin}\tikzstyle{every node}=[font=\tiny]
    \node[align=left, scale=0.5] at (\dynkinstep*1,-.15cm){1};
    \node[align=left, scale=0.5] at (\dynkinstep*2,-.15cm){3};
    \node[align=left, scale=0.5] at (\dynkinstep*3,-.15cm){4};
    \node[align=left, scale=0.5] at (\dynkinstep*4,-.15cm){5};
    \node[align=left, scale=0.5] at (\dynkinstep*5,-.15cm){6};
    \node[align=left, scale=0.5] at (\dynkinstep*6,-.15cm){7};
    \node[align=left, scale=0.5] at (\dynkinstep*3.5,.25cm){2};
    \foreach \x in {1,...,6}
    {
        \dynkindot{\x}{0}
    }
    \dynkindot{3}{1}
    \dynkinline{1}{0}{6}{0}
    \dynkinline{3}{0}{3}{1}
  \end{dynkin}}& $(2,2,3,4,3,2,1)$  & $(0,0,0,0,0,0,1)$ & 12\\
\hline 
$E_8$ & \scalebox{3}{\begin{dynkin}\tikzstyle{every node}=[font=\tiny]
    \node[align=left, scale=0.5] at (\dynkinstep*1,-.15cm){1};
    \node[align=left, scale=0.5] at (\dynkinstep*2,-.15cm){3};
    \node[align=left, scale=0.5] at (\dynkinstep*3,-.15cm){4};
    \node[align=left, scale=0.5] at (\dynkinstep*4,-.15cm){5};
    \node[align=left, scale=0.5] at (\dynkinstep*5,-.15cm){6};
    \node[align=left, scale=0.5] at (\dynkinstep*6,-.15cm){7};
    \node[align=left, scale=0.5] at (\dynkinstep*7,-.15cm){8};
    \node[align=left, scale=0.5] at (\dynkinstep*3.5,.25cm){2};
    \foreach \x in {1,...,7}
    {
        \dynkindot{\x}{0}
    }
    \dynkindot{3}{1}
    \dynkinline{1}{0}{7}{0}
    \dynkinline{3}{0}{3}{1}
  \end{dynkin}}& $(2,3,4,6,5,4,3,2)$  & $(0,0,0,0,0,0,0,1)$ & 60\\
\hline 
$F_4$ &  \scalebox{3}{\begin{dynkin}\tikzstyle{every node}=[font=\tiny]
    \node[align=left, scale=0.5] at (\dynkinstep*1,-.15cm){1};
    \node[align=left, scale=0.5] at (\dynkinstep*2,-.15cm){2};
    \node[align=left, scale=0.5] at (\dynkinstep*3,-.15cm){3};
    \node[align=left, scale=0.5] at (\dynkinstep*4,-.15cm){4};
    \node[align=left, scale=0.5] at (\dynkinstep*1,.1cm){};
    \dynkindoubleline{2}{0}{3}{0}
    \foreach \x in {1,...,4}
    {
        \dynkindot{\x}{0}
    }
    \dynkinline{1}{0}{2}{0}
    \dynkinline{3}{0}{4}{0}
  \end{dynkin}}& $(2,3,4,2)$   & $(0,0,0,1)$ & 6\\
\hline 
$G_2$ &  \scalebox{3}{\begin{dynkin}\tikzstyle{every node}=[font=\tiny]
    \node[align=left, scale=0.5] at (\dynkinstep*1,-.15cm){1};
    \node[align=left, scale=0.5] at (\dynkinstep*2,-.15cm){2};
    \node[align=left, scale=0.5] at (\dynkinstep*1,.1cm){};
    \dynkintripleline{2}{0}{1}{0}
    \foreach \x in {1,2}
    {
        \dynkindot{\x}{0}
    }
  \end{dynkin}} & $(3,2)$  & $(1,0)$ & 2 \\
\hline
\end{tabular}
\end{center}
\caption{Conventions for the simple Lie algebras.}
\label{t:lie_conventions}
\end{table}

\newpage

\bibliography{references}
\bibliographystyle{utcaps}

\end{document}